\documentclass[aps,prc,amsmath,amssymb,showpacs,superscriptaddress,nofootinbib,10pt]{revtex4-2}
\usepackage{graphicx}% Include figure files
\usepackage{dcolumn}% Align table columns on decimal point
\usepackage{bm}% bold math
\usepackage[utf8]{inputenc}
\usepackage{color}
\usepackage{orcidlink}
\usepackage{hyperref}% add hypertext capabilities
\usepackage{verbatim}
\usepackage{tabularx}
\usepackage{booktabs}
\usepackage{array}
\usepackage{braket}
\newcolumntype{C}[1]{>{\centering\let\newline\\\arraybackslash\hspace{0pt}}m{#1}}
\usepackage{multirow}
\usepackage{mathtools}
\allowdisplaybreaks
\newcommand{\fett}[1]{\mbox{\boldmath $#1$}}

\newcommand{\beq}{\begin{equation}}
	\newcommand{\eeq}{\end{equation}}
\newcommand{\beqa}{\begin{eqnarray}}
	\newcommand{\eeqa}{\end{eqnarray}}
\newcommand{\nn}{\nonumber \\ }

\newcommand{\RN}[1]{%
  \textup{\uppercase\expandafter{\romannumeral#1}}%
}

\DeclareFontEncoding{LS1}{}{}
\DeclareFontSubstitution{LS1}{stix}{m}{n}
\DeclareSymbolFont{stixletters}{LS1}{stix}{m}{it}
\DeclareMathAccent{\cev}{\mathord}{stixletters}{"91}
\DeclareMathAccent{\vec}{\mathord}{stixletters}{"92}
\DeclareMathAccent{\vecev}{\mathord}{stixletters}{"95}

\begin{document}

\title{Subleading \textit{D}-like Three-Nucleon Interactions}% Force line breaks with \\
\author{Henri Paul Huesmann\orcidlink{0009-0006-2092-961X}}
\affiliation{Ruhr-Universit\"at Bochum, Fakult\"at f\"ur Physik und Astronomie, Institut f\"ur Theoretische Physik II, D-44780 Bochum, Germany}

\author{Hermann Krebs\orcidlink{0000-0002-2263-0308}}
\affiliation{Ruhr-Universit\"at Bochum, Fakult\"at f\"ur Physik und Astronomie, Institut f\"ur Theoretische Physik II, D-44780 Bochum, Germany}

\author{Evgeny Epelbaum\orcidlink{0000-0002-7613-0210}}
\affiliation{Ruhr-Universit\"at Bochum, Fakult\"at f\"ur Physik und Astronomie, Institut f\"ur Theoretische Physik II, D-44780 Bochum, Germany}

\begin{abstract}
  We consider subleading contributions to the three-nucleon force from tree-level diagrams involving a single-pion exchange and a contact interaction between two nucleons, which appear at fifth order in the chiral expansion. We show that the corresponding $D$-like three-nucleon potential depends on $16$ low-energy constants, which need to be determined from few-body data. Assuming that their numerical values are governed by the intermediate $\Delta$(1232) excitation mechanism, the
considered three-nucleon force can be approximated using $4$ low-energy constants that parametrize the short-range nucleon-nucleon to nucleon-$\Delta$ transition amplitude. 
\end{abstract}

\maketitle

%\textit{Introduction}---
\section{Introduction}

Three-body interactions are known to play an important role in atomic and nuclear physics \cite{Endo:2024cbz}. Compared with two-nucleon (NN) interactions, which have been extensively studied in the framework of chiral effective field theory (EFT) \cite{Epelbaum:2008ga,Machleidt:2011zz,Epelbaum:2019kcf} and can be regarded as well-established, three-nucleon forces (3NFs) remain poorly understood. They are widely viewed as a critical missing ingredient for advancing {\it ab-initio} description of nuclear structure and the equation of state of neutron/nuclear matter \cite{Hebeler:2009iv}, thus constituting an important frontier in nuclear physics \cite{Kalantar-Nayestanaki:2011rzs,Hammer:2012id,Endo:2024cbz}. 

In the two-nucleon sector of chiral EFT, the expansion of the interaction has been pushed to (and even beyond) the order $Q^5$ (i.e., N$^4$LO), where $Q \in \{M_\pi/\Lambda_b, \, |\vec p \, |/\Lambda_b \}$ denotes the expansion parameter with $M_\pi$, $\Lambda_b$ and $\vec p$ referring to the pion mass, the EFT breakdown scale and characteristic external momenta of the nucleons, respectively. The resulting N$^4$LO$^+$ potentials or Refs.~\cite{Reinert:2017usi,Reinert:2020mcu} lead to a statistically perfect description of mutually compatible neutron-proton and proton-proton scattering data below pion production threshold \cite{Reinert:2022jpu,Epelbaum:2022cyo}. Three-nucleon interactions start contributing at order  $Q^3$ (i.e., N$^2$LO) from tree-level diagrams of the two-pion exchange, one-pion-contact ($1\pi$-contact) and purely short-range types \cite{vanKolck:1994yi,Epelbaum:2002vt}. Since the relevant pion-nucleon ($\pi$N) low-energy constants (LECs) can be reliably determined from experimental data on the $\pi$N system \cite{Hoferichter:2015tha,Siemens:2016jwj}, the long-range part of the 3NF comes out as a parameter-free prediction. On the other hand, the leading short-range contributions to the 3NF depend on the LECs $D$ and $E$, which need to be determined from experimental data on $A\ge 3$ systems.  Subleading contributions to the 3NF at order $Q^4$ (i.e., N$^3$LO) arise from loop diagrams constructed from the lowest-order vertices in the effective Lagrangian and have been derived in Refs.~\cite{Ishikawa:2007zz,Bernard:2007sp,Bernard:2011zr}. Furthermore, N$^4$LO corrections to the long-range two-pion exchange, one-pion-two-pion-exchange and ring topologies have been worked out in Refs.~\cite{Krebs:2012yv,Krebs:2013kha}, while the subleading contributions to the purely contact 3NF are discussed in Ref.~\cite{Girlanda:2011fh} and depend on $13$ unknown LECs $E_i$. Selected types of 3NFs at order $Q^6$ (N$^5$LO) are discussed in Refs.~\cite{Cirigliano:2024ocg,Epelbaum:2025wtj}, while $\Delta$-isobar contributions are considered in Refs.~\cite{Epelbaum:2007sq,Krebs:2018jkc}. 

In spite of these developments, the current accuracy level of chiral EFT studies beyond the NN system is limited to N$^2$LO, see Ref.~\cite{Epelbaum:2019zqc,Maris:2020qne,LENPIC:2022cyu} for applications to nucleon-deuteron scattering, with the main obstacle being the lack of consistently regularized 3NFs: It is shown in Ref.~\cite{Epelbaum:2019kcf} that imposing a cutoff regulator on three-nucleon potentials derived using dimensional regularization \cite{Bernard:2007sp,Bernard:2011zr,Krebs:2012yv,Krebs:2013kha} leads to violation of chiral symmetry. The same issue affects loop contributions to the exchange current operators \cite{Krebs:2020pii}. As a consequence, loop contributions to the 3NF and exchange current operators need to be re-derived using a symmetry-preserving cutoff regulator. A suitable theoretical framework to perform such calculations has been established only recently based on the gradient flow method \cite{Krebs:2023ljo,Krebs:2023gge}, and the gradient-flow-regularized contributions to the 3NF at N$^3$LO are currently under investigation. In the meantime, it is both timely and important to address the question of constraining the short-range part of the 3NF, which is considered to be a significant computational task. Exploratory studies of three-nucleon scattering observables using the purely short-range subleading three-nucleon interactions at N$^4$LO have revealed promising results by showing that the inclusion of the $E_i$-terms is potentially capable of resolving some of the long-standing discrepancies between theory and experimental data \cite{Girlanda:2018xrw,Epelbaum:2019zqc,Witala:2022rzl,Girlanda:2023znc}. On the other hand, subleading $1\pi$-contact three-nucleon interactions have not been investigated so far, and even the number of LECs needed to parametrize these 3NF contributions at N$^4$LO remains unknown.

In this paper we fill this gap by analyzing tree-level corrections to the \textit{D}-like 3NFs. Our paper is organized as follows. In sec.~\ref{sec:DLike3NF}, we use three different methods to work out the minimal set of operators contributing to the considered type of 3NF at N$^4$LO and show that it can be parametrized in terms of $16$ LECs $F_{1, \ldots , 16}$. In sec.~\ref{sec:3NFIsospin}, we analyze the isospin structure of the resulting \textit{D}-like 3NFs to isolate possible combinations of the LECs that do not contribute to nucleon-deuteron scattering in the isospin symmetry limit. Next, the interplay between the off-shell part of the short-range NN force and the considered type of 3NFs and implications of making specific choices for the off-shell behavior of the NN force as done in Refs.~\cite{Reinert:2017usi,Reinert:2020mcu} are discussed in sec.~\ref{sec:UT}. In sec.~\ref{sec:Delta}, we address the role of the intermediate $\Delta$(1232) excitations and estimate the dominant contributions to $F_i$'s by using the resonance saturation hypothesis. Finally, the main results of this paper are summarized in sec.~\ref{sec:summary}.

\section{Construction of the \textit{D}-like three-nucleon operators at N$^4$LO}
\label{sec:DLike3NF}

  As already mentioned above, the leading contribution to the $1\pi$-contact 3NF appears at N$^2$LO and depends on the single LEC $D$ \cite{vanKolck:1994yi,Epelbaum:2002vt}. Tree-level corrections of the $1\pi$-contact type at N$^3$LO can arise either from subleading $\pi NN$ vertices involving two derivatives or from the leading relativistic corrections to reducible-like NLO diagrams, whose 3NF contributions vanish in the static limit \cite{vanKolck:1994yi,Epelbaum:1998ka}. The latter type of contributions $\propto g_A^2 C_{S,T} /m$, where $C_{S,T}$ denote the LECs parametrizing the leading-order NN contact interactions \cite{Weinberg:1990rz,Weinberg:1991um},  are given in Ref.~\cite{Bernard:2011zr}. The former type of 3NF contributions can only appear beyond N$^4$LO in the counting scheme for the nucleon mass $m$ with $m \sim \Lambda_b^2/M_\pi \gg \Lambda_b$ \cite{Weinberg:1991um,Epelbaum:2008ga,Epelbaum:2019kcf}, which we employ here. This can be understood by observing that $\pi$NN vertices with two spatial derivatives are forbidden by parity conservation. On the other hand, $\pi$NN vertices with one spatial and one time derivative are possible, but the corresponding  tree-level three-nucleon-irreducible diagrams acquire an additional suppression factor of $\sim p/m$. Accordingly, the resulting  3NFs are suppressed at least by a factor of $p^2/(m\Lambda_b)$ relative to the N$^2$LO 3NF and, therefore, start contributing beyond N$^4$LO.

In the following, we focus on the tree-level contributions to the $1\pi$-contact 3NF at N$^4$LO, which emerge from $\pi$NN vertices involving three spatial derivatives, and work out a minimal set of independent structures. For the sake of validation, we use three different approaches: First we directly parametrize the most general form of the 3NF of the required type in terms of relative momenta of the nucleons. In the second approach, we proceed by constructing the non-relativistic Lagrangian with four nucleon fields and a single pion field involving three derivatives and make use of reparametrization invariance to account for constraints provided by Poincar\'e symmetry. In the last considered approach, we start from the covariant Lagrangian and perform its non-relativistic reduction. As will be shown below, all approaches lead to the same result and show that the momentum dependence of the subleading N$^4$LO $D$-like 3NFs is parametrizable in terms of $16$ LECs. 

\subsection{Method I: Direct parametrization of the potential in terms of relative momenta}
\label{sec:Method1}

To derive the expression for the subleading tree-level $D$-like 3NF, one can proceed by parametrizing the most general form of the short-range NN$\, \to \,$NN$\pi$ transition amplitude, which is compatible with the required symmetry principles to be specified below and the expansion order of chiral EFT. Specifically, we aim at constructing all possible two-nucleon operators $\Tilde{\fett O} \sim \mathcal{O}(Q^3)$ of isovector type, which are made out of the momenta $\Vec{q}$, $\Vec{k}$ and $\Vec{q_3}$, Pauli spin matrices $\Vec{\sigma}_1$ and  $\Vec{\sigma}_2$ of nucleons $1$ and $2$ as well as the corresponding isospin matrices $\boldsymbol{\tau}_1$ and $\boldsymbol{\tau}_2$. Here, the momenta $\Vec{q}$ and $\Vec{k}$ are defined in terms of the relative incoming (outgoing) nucleon momentum $\vec p = (\Vec{p}_1 - \Vec{p}_2)/2$ ($\Vec{p}' = (\Vec{p}'_1 - \Vec{p}'_2)/2$) via $\Vec{q} = \Vec{p}' - \Vec{p}$ and $\Vec{k} = (\Vec{p}' + \Vec{p})/2$, while $\Vec{q_3} = \Vec{p}'_3 - \Vec{p}_3$ denotes the momentum of the virtual pion exchanged between the nucleon $3$ and the subsystem $(12)$. The reliance on the relative momenta ensures Galilean invariance of the resulting 3N potential. Terms depending on the total momentum are expected to appear in the form of relativistic corrections suppressed by inverse powers of the nucleon mass $m$, see sec.~\ref{sec:Method3} for a discussion. In the employed counting scheme for the nucleon mass, such relativistic corrections appear beyond N$^4$LO and are not considered here.   
Notice further that the assumed dependence of $\Tilde{\fett O}$ on only three out of four Jacobi momenta is motivated by the local nature of the one-pion exchange.

The corresponding contributions to the 3NF at order N$^4$LO can be obtained by combining the operators $\Tilde{\fett O}$ with the pion propagator and the pion-nucleon vertex via
\beq
\label{PotFromO}
V \propto \frac{\Tilde{\fett O} \cdot \fett \tau_3 \, \Vec{\sigma_3} \cdot \Vec{q_3}}{q_3^2 + M_{\pi}^2}\,.
\eeq
Notice that one, in principle, also needs to consider contributions stemming from higher-order corrections to the lowest-order $\pi$N vertex proportional to the axial coupling constant of the nucleon $g_A$. The leading relativistic corrections to the $g_A$-vertex involve a time derivative of the pion field and are suppressed by the factor of $1/m$. For irreducible tree-level diagrams, the time derivative of the pion field leads to the energy transfer of the nucleons and thus results in an additional suppression factor of $\sim p/m$. Accordingly, the corresponding relativistic corrections are suppressed by a factor of $p^2/m^2 \sim \mathcal{O} (p^4/\Lambda_b^4)$. Even for the N$^2$LO $1\pi$-contact 3NF $\propto D$, such relativistic corrections appear well beyond the N$^4$LO accuracy level of this study. The remaining corrections to the $\pi$N vertex $\propto g_A$ involve either insertions of  $M_\pi^2$ or higher powers of the pion momentum $\vec q_3$ and thus can be included by replacing $g_A$ with its effective value to account for the Goldberger-Treiman discrepancy and by shifting the values of the LECs $D$ and $E_i$. We thus do not consider such type of corrections and focus solely on the operators $\Tilde{\fett O}_i$ of the order $Q^3$.

Specifically, we are looking for operators $\Tilde{\fett O}_i$ which are parity ($\mathcal{P}$) odd, rotationally invariant and symmetric with respect to permutations of the nucleons $1 \leftrightarrow 2$.
Furthermore, we demand that each Levi-Civita symbol appears together with a factor of $i$ to ensure that the resulting 3NF is time-reversal invariant. 
% and time-reversal ($\mathcal{T}$) invariant.
% ($\mathcal{C}-$invariance is thus fulfilled by CPT),
% The last requirement is achieved by demanding that each Levi-Civita symbol appears together with a factor of $i$.
Finally, we require the operators $\Tilde{\fett O}_i$ to be odd under hermitian conjugation in order that $V$ in Eq.~(\ref{PotFromO}) is hermitian. This implies that the operators $\Tilde{\fett O}_i$ involving a factor of $i$ must be proportional to $\vec k$, while those without a factor of $i$ are either independent or quadratic in $\vec k$. Notice that each operator needs to involve at least one power of the pion momentum $\vec q_3$ to comply with the Goldstone-boson nature of pions.  
\begin{table*}
				\begin{ruledtabular}
					\begin{tabular*}{\textwidth}{@{\extracolsep{\fill}}llll}
\phantom{xx}$\Tilde{\fett O}_1$ & $ i \; (\Vec{q} \times \Vec{k} ) \cdot \Vec{q_3} \; (\pmb{\tau}_1 + \pmb{\tau}_2)$ \phantom{xxxxxxxxxxxxxxxxxxxxxxxx} & $\Tilde{\fett O}_{22}$ & $ q^2 \; (\Vec{\sigma}_1 \times \Vec{\sigma}_2) \cdot \Vec{q_3}  \; (\pmb{\tau}_1 \times \pmb{\tau}_2) $ \\

\phantom{xx}$\Tilde{\fett O}_2$  & $ (\Vec{\sigma}_1 + \Vec{\sigma}_2) \cdot \Vec{q} \;\; \Vec{q}  \cdot \Vec{q_3} \; (\pmb{\tau}_1 + \pmb{\tau}_2) $ & $\Tilde{\fett O}_{23}$ & $   \Vec{\sigma}_1 \cdot \Vec{q} \; (\Vec{q} \times \Vec{\sigma}_2) \cdot \Vec{q_3}  \; (\pmb{\tau}_1 \times \pmb{\tau}_2)  + 1 \leftrightarrow 2 $ \\

\phantom{xx}$\Tilde{\fett O}_3$  & $ q^2 \; (\Vec{\sigma}_1 + \Vec{\sigma}_2) \cdot \Vec{q_3} \; (\pmb{\tau}_1 + \pmb{\tau}_2) $ & $\Tilde{\fett O}_{24}$ & $ i (\Vec{\sigma}_1 \times \Vec{\sigma}_2) \cdot \Vec{q} \;\; \Vec{k}  \cdot \Vec{q_3} \; (\pmb{\tau}_1 - \pmb{\tau}_2) $ \\

\phantom{xx}$\Tilde{\fett O}_4$  & $ (\Vec{\sigma}_1 - \Vec{\sigma}_2) \cdot \Vec{q} \;\; \Vec{q}  \cdot \Vec{q_3} \; (\pmb{\tau}_1 - \pmb{\tau}_2) $ & $\Tilde{\fett O}_{25}$ & $ i (\Vec{\sigma}_1 \times \Vec{\sigma}_2) \cdot \Vec{k} \;\; \Vec{q}  \cdot \Vec{q_3} \; (\pmb{\tau}_1 - \pmb{\tau}_2) $ \\

\phantom{xx}$\Tilde{\fett O}_5$  & $ q^2 \; (\Vec{\sigma}_1 - \Vec{\sigma}_2) \cdot \Vec{q_3} \; (\pmb{\tau}_1 - \pmb{\tau}_2) $ & $\Tilde{\fett O}_{26}$ & $ i \Vec{k}  \cdot \Vec{q} \;\; (\Vec{\sigma}_1 \times \Vec{\sigma}_2) \cdot \Vec{q_3}  \; (\pmb{\tau}_1 - \pmb{\tau}_2) $ \\

\phantom{xx}$\Tilde{\fett O}_6$  & $ i (\Vec{\sigma}_1 - \Vec{\sigma}_2) \cdot \Vec{q} \;\; \Vec{k}  \cdot \Vec{q_3} \; (\pmb{\tau}_1 \times \pmb{\tau}_2) $ & $\Tilde{\fett O}_{27}$ & $ i (\Vec{k}  \times \Vec{q}) \cdot \Vec{q_3} \; \Vec{\sigma}_1 \cdot \Vec{\sigma}_2  \; (\pmb{\tau}_1 + \pmb{\tau}_2) $ \\

\phantom{xx}$\Tilde{\fett O}_7$  & $ i (\Vec{\sigma}_1 - \Vec{\sigma}_2) \cdot \Vec{k} \;\; \Vec{q}  \cdot \Vec{q_3} \; (\pmb{\tau}_1 \times \pmb{\tau}_2) $ & $\Tilde{\fett O}_{28}$ & $  i \Vec{\sigma}_2 \cdot \Vec{q} \; (\Vec{\sigma}_1 \times \Vec{k}) \cdot \Vec{q_3}  \; (\pmb{\tau}_1 + \pmb{\tau}_2)  + 1 \leftrightarrow 2 $ \\

\phantom{xx}$\Tilde{\fett O}_8$  & $ i \Vec{k}  \cdot \Vec{q} \;\;  (\Vec{\sigma}_1 - \Vec{\sigma}_2) \cdot \Vec{q_3} \;  (\pmb{\tau}_1 \times \pmb{\tau}_2) $ & $\Tilde{\fett O}_{29}$ & $ i  \Vec{\sigma}_2 \cdot \Vec{k} \; (\Vec{q} \times \Vec{\sigma}_1) \cdot \Vec{q_3}  \; (\pmb{\tau}_1 + \pmb{\tau}_2)  + 1 \leftrightarrow 2 $ \\

\phantom{xx}$\Tilde{\fett O}_9$  & $ (\Vec{\sigma}_1 + \Vec{\sigma}_2) \cdot \Vec{q} \; \; q_3^2 \; (\pmb{\tau}_1 - \pmb{\tau}_2) $ & $\Tilde{\fett O}_{30}$ & $ i \; (\Vec{q} \times \Vec{\sigma}_1) \cdot \Vec{k} \;\;  \Vec{\sigma}_2 \cdot \Vec{q_3} \; (\pmb{\tau}_1 + \pmb{\tau}_2)  + 1 \leftrightarrow 2 $ \phantom{xx}\\

\phantom{xx}$\Tilde{\fett O}_{10}$  & $ (\Vec{\sigma}_1 + \Vec{\sigma}_2) \cdot \Vec{q_3} \; \; \Vec{q}  \cdot \Vec{q_3} \; (\pmb{\tau}_1 - \pmb{\tau}_2) $ & $\Tilde{\fett O}_{31}$ & $ i \Vec{\sigma}_1 \cdot \Vec{q} \; (\Vec{\sigma}_2 \times \Vec{k}) \cdot \Vec{q_3}  \;\; (\pmb{\tau}_1 - \pmb{\tau}_2)  + 1 \leftrightarrow 2 $ \\

\phantom{xx}$\Tilde{\fett O}_{11}$  & $ (\Vec{\sigma}_1 - \Vec{\sigma}_2) \cdot \Vec{q} \; \; q_3^2 \; (\pmb{\tau}_1 + \pmb{\tau}_2) $ & $\Tilde{\fett O}_{32}$ & $ i \Vec{\sigma}_2 \cdot \Vec{k} \; ( \Vec{q} \times  \Vec{\sigma}_1) \cdot \Vec{q_3}  \;\; (\pmb{\tau}_1 - \pmb{\tau}_2)  + 1 \leftrightarrow 2 $ \\

\phantom{xx}$\Tilde{\fett O}_{12}$  & $ (\Vec{\sigma}_1 - \Vec{\sigma}_2) \cdot \Vec{q_3} \; \; \Vec{q}  \cdot \Vec{q_3} \; (\pmb{\tau}_1 + \pmb{\tau}_2) $ & $\Tilde{\fett O}_{33}$ & $ i \; (\Vec{q} \times \Vec{\sigma}_1) \cdot \Vec{k} \;\;  \Vec{\sigma}_2 \cdot \Vec{q_3} \; (\pmb{\tau}_1 - \pmb{\tau}_2)  + 1 \leftrightarrow 2 $ \\

\phantom{xx}$\Tilde{\fett O}_{13}$  & $ (\Vec{\sigma}_1 + \Vec{\sigma}_2) \cdot \Vec{k} \;\; \Vec{k}  \cdot \Vec{q_3} \; (\pmb{\tau}_1 + \pmb{\tau}_2) $ & $\Tilde{\fett O}_{34}$ & $ i \Vec{\sigma}_1 \cdot \Vec{q_3} \; ( \Vec{q} \times  \Vec{\sigma}_2) \cdot \Vec{q_3}  \;\; (\pmb{\tau}_1 \times \pmb{\tau}_2)  + 1 \leftrightarrow 2 $ \\

\phantom{xx}$\Tilde{\fett O}_{14}$  & $ k^2 \; (\Vec{\sigma}_1 + \Vec{\sigma}_2) \cdot \Vec{q_3} \; \; (\pmb{\tau}_1 + \pmb{\tau}_2) $ & $\Tilde{\fett O}_{35}$  & $  (\Vec{\sigma}_1 \times \Vec{\sigma}_2) \cdot \Vec{k} \;\; \Vec{k}  \cdot \Vec{q_3} \; (\pmb{\tau}_1 \times \pmb{\tau}_2) $ \\

\phantom{xx}$\Tilde{\fett O}_{15}$  & $ (\Vec{\sigma}_1 - \Vec{\sigma}_2) \cdot \Vec{k} \;\; \Vec{k}  \cdot \Vec{q_3} \; (\pmb{\tau}_1 - \pmb{\tau}_2) $& $\Tilde{\fett O}_{36}$  & $  k^2 \; (\Vec{\sigma}_1 \times \Vec{\sigma}_2) \cdot \Vec{q_3}  \; (\pmb{\tau}_1 \times \pmb{\tau}_2) $ \\

\phantom{xx}$\Tilde{\fett O}_{16}$  & $ k^2 \; (\Vec{\sigma}_1 - \Vec{\sigma}_2) \cdot \Vec{q_3} \; \; (\pmb{\tau}_1 - \pmb{\tau}_2) $ & $\Tilde{\fett O}_{37}$ & $  \Vec{\sigma}_2 \cdot \Vec{k} \; ( \Vec{\sigma}_1 \times \Vec{k}) \cdot \Vec{q_3}  \;\; (\pmb{\tau}_1 \times \pmb{\tau}_2)  + 1 \leftrightarrow 2 $ \\

\phantom{xx}$\Tilde{\fett O}_{17}$  & $ i (\Vec{\sigma}_1 + \Vec{\sigma}_2) \cdot \Vec{k} \;\; q_3^2 \; (\pmb{\tau}_1 \times \pmb{\tau}_2) $ & $\Tilde{\fett O}_{38}$ & $ i (\Vec{\sigma}_1 \times \Vec{\sigma}_2) \cdot \Vec{k} \;\; q_3^2 \; (\pmb{\tau}_1 + \pmb{\tau}_2) $ \\

\phantom{xx}$\Tilde{\fett O}_{18}$  & $ i (\Vec{\sigma}_1 + \Vec{\sigma}_2) \cdot \Vec{q_3} \;\; \Vec{k}  \cdot \Vec{q_3} \; (\pmb{\tau}_1 \times \pmb{\tau}_2) $ & $\Tilde{\fett O}_{39}$ & $ i (\Vec{\sigma}_1 \times \Vec{\sigma}_2) \cdot \Vec{q_3} \;\; \Vec{k} \cdot \Vec{q_3} \; (\pmb{\tau}_1 + \pmb{\tau}_2) $ \\

\phantom{xx}$\Tilde{\fett O}_{19}$  & $ (\Vec{\sigma}_1 + \Vec{\sigma}_2) \cdot \Vec{q_3} \;\; q_3^2 \; (\pmb{\tau}_1 + \pmb{\tau}_2) $ & $\Tilde{\fett O}_{40}$ & $ i \Vec{\sigma}_2 \cdot \Vec{q_3} \; ( \Vec{\sigma}_1 \times \Vec{k}) \cdot \Vec{q_3}  \;\; (\pmb{\tau}_1 - \pmb{\tau}_2)  + 1 \leftrightarrow 2 $ \\

\phantom{xx}$\Tilde{\fett O}_{20}$  & $ (\Vec{\sigma}_1 - \Vec{\sigma}_2) \cdot \Vec{q_3} \;\; q_3^2 \; (\pmb{\tau}_1 - \pmb{\tau}_2) $ & $\Tilde{\fett O}_{41}$ & $ i  \Vec{\sigma}_1 \cdot \Vec{q_3} \; ( \Vec{\sigma}_2 \times \Vec{k}) \cdot \Vec{q_3}  \;\; (\pmb{\tau}_1 + \pmb{\tau}_2)  + 1 \leftrightarrow 2 $ \\

\phantom{xx}$\Tilde{\fett O}_{21}$  & $ (\Vec{\sigma}_1 \times \Vec{\sigma}_2) \cdot \Vec{q} \;\; \Vec{q}  \cdot \Vec{q_3} \; (\pmb{\tau}_1 \times \pmb{\tau}_2) $ & $\Tilde{\fett O}_{42}$ & $   (\Vec{\sigma}_1 \times \Vec{\sigma}_2) \cdot \Vec{q_3} \;\; q_3^2 \; (\pmb{\tau}_1 \times \pmb{\tau}_2) $ 
                                        \end{tabular*}
 \caption{The complete list of $42$ two-nucleon operators $\Tilde{\fett O}_i$ needed to construct the $D$-like 3NF at N$^4$LO.}
\label{Table1}                                    
  \end{ruledtabular}
\end{table*}

We found $42$ operators $\Tilde{\fett O}_i$ that fulfill all above requirements, see Table \ref{Table1}. 
However, these operators are not all independent since we have not yet taken into account constraints imposed by the Pauli principle. To antisymmetrize the potentials with respect to nucleons $1$ and $2$, we act on $\Tilde{\fett O}_i$ with the antisymmetrization operator $\mathcal{A}_{12} = \frac{1-P_{12}}{2}$, where $P_{12}$ denotes the permutation operator, $P_{12} \ket{ij} = \ket{ji}$. Its spin-isospin part can be written in terms of the corresponding exchange operators,
\beq
P^{\rm spin-isospin}_{12} = \frac{1+\vec{\sigma}_1 \cdot \vec{\sigma}_2 }{2} \, \frac{1+{\fett \tau}_1 \cdot {\fett \tau}_2 }{2},
\eeq
while the momentum part amounts to interchanging the nucleon momenta, i.e.~$\Vec{p}' \to -\Vec{p}'$ if the antisymmetrizer is applied on the final state. Performing antisymmetrization of the operators $\Tilde{\fett O}_i$ and making use of Schouten identities, we find the following relationships between the operators:\footnote{Here and in what follows, algebraic calculations are performed using the software {\it Mathematica} \cite{Mathematica}.} 
\begin{eqnarray}
  \Tilde{\fett O}_{1} &=& \Tilde{\fett O}_{27} = \Tilde{\fett O}_{28} - \frac{1}{2}\Tilde{\fett O}_{30}, \nn
     \Tilde{\fett O}_{2} &= &- 4  \Tilde{\fett O}_{13}, \nn
                \Tilde{\fett O}_{3} &=& -4  \Tilde{\fett O}_{14}, \nn
                          \Tilde{\fett O}_{4} &=& 4 \Tilde{\fett O}_{35} =  \Tilde{\fett O}_{5} - 4 \Tilde{\fett O}_{37}, \nn
                                    \Tilde{\fett O}_{5} &=& 4 \Tilde{\fett O}_{36}, \nn
                                              \Tilde{\fett O}_{6} &=& - \Tilde{\fett O}_{25} = \Tilde{\fett O}_{8} - \Tilde{\fett O}_{31} = \Tilde{\fett O}_{7} + \Tilde{\fett O}_{33}, \nn
                                                        \Tilde{\fett O}_{7} &=& - \Tilde{\fett O}_{24} = \Tilde{\fett O}_{8} - \Tilde{\fett O}_{32}, \nn
                                                                  \Tilde{\fett O}_{8} &=& - \Tilde{\fett O}_{26}, \nn
                                                        \Tilde{\fett O}_{9} &=& 2 \Tilde{\fett O}_{17}, \nn
                                                                  \Tilde{\fett O}_{10} &=& 2 \Tilde{\fett O}_{18},\nn
                                                                             \Tilde{\fett O}_{11} &=& 2 \Tilde{\fett O}_{38} =  \Tilde{\fett O}_{12} + 2 \Tilde{\fett O}_{41}, \nn
                                                                                        \Tilde{\fett O}_{12} &=& 2 \Tilde{\fett O}_{39}, \nn
                                                                                                   \Tilde{\fett O}_{15} &=& \frac{1}{4}\Tilde{\fett O}_{21} = \Tilde{\fett O}_{16} -  \frac{1}{4}\Tilde{\fett O}_{23}, \nn
                                                                                                                \Tilde{\fett O}_{16} &=& \frac{1}{4}\Tilde{\fett O}_{22}, \nn
                                                                                                                             \Tilde{\fett O}_{19} &=& 0, \nn
                                                                                                                                        \Tilde{\fett O}_{20} &=& \Tilde{\fett O}_{42}, \nn
                                                                                                                                                   \Tilde{\fett O}_{28} &=& \Tilde{\fett O}_{29}, \nn
                                                                                                                                                              \Tilde{\fett O}_{34} &=& 2 \Tilde{\fett O}_{40}.
 \label{Eq1}
\end{eqnarray}
Accordingly, we end up with $17$ independent operators, namely $11$ operators proportional to $\vec q_3$, $5$ operators quadratic in  $\vec q_3$ and one operator cubic  in  $\vec q_3$.
Clearly, the choice of the operator basis prior to antisymmetrization is not unique. In Table \ref{Table2}, we give a possible choice of independent operators in which we kept as many terms local, i.e., depending only on the momenta $\Vec{q}$ and $\Vec{q}_3$, as possible. We further emphasize that by expressing $q_3^2$ in terms of the inverse pion propagator, the 3NF corresponding to the operator ${\fett O}_{17}$ can be absorbed into the appropriate shifts of the LECs $D$ and $E_i$. 
\begin{table*}
				\begin{ruledtabular}
					\begin{tabular*}{\textwidth}{@{\extracolsep{\fill}}llll}
\phantom{xx}${\fett O}_1$ & $ i \; (\Vec{q} \times \Vec{k} ) \cdot \Vec{q_3} \; (\pmb{\tau}_1 + \pmb{\tau}_2) $ \phantom{xxxxxxxxxxxxxxxxxxxxxxxxxxxxxx} & ${\fett O}_{10}$ \phantom{xx}  & $ (\Vec{\sigma}_1 + \Vec{\sigma}_2) \cdot \Vec{q} \;\; \Vec{q}  \cdot \Vec{q_3} \; (\pmb{\tau}_1 + \pmb{\tau}_2) $ \\

\phantom{xx}${\fett O}_2$  & $ i (\Vec{\sigma}_1 - \Vec{\sigma}_2) \cdot \Vec{k} \;\; \Vec{q}  \cdot \Vec{q_3} \; (\pmb{\tau}_1 \times \pmb{\tau}_2) $ & ${\fett O}_{11}$ & $  i \Vec{\sigma}_2 \cdot \Vec{q} \; (\Vec{\sigma}_1 \times \Vec{k}) \cdot \Vec{q_3}  \; (\pmb{\tau}_1 + \pmb{\tau}_2)  + 1 \leftrightarrow 2 $ \\

\phantom{xx}${\fett O}_3$  & $ i \Vec{k}  \cdot \Vec{q} \;\;  (\Vec{\sigma}_1 - \Vec{\sigma}_2) \cdot \Vec{q_3} \;  (\pmb{\tau}_1 \times \pmb{\tau}_2) $ & ${\fett O}_{12}$ & $ i \Vec{\sigma}_1 \cdot \Vec{q_3} \; ( \Vec{q} \times  \Vec{\sigma}_2) \cdot \Vec{q_3}  \;\; (\pmb{\tau}_1 \times \pmb{\tau}_2)  + 1 \leftrightarrow 2 $ \\

\phantom{xx}${\fett O}_4$  & $ i (\Vec{\sigma}_1 - \Vec{\sigma}_2) \cdot \Vec{q} \;\; \Vec{k}  \cdot \Vec{q_3} \; (\pmb{\tau}_1 \times \pmb{\tau}_2) $ & ${\fett O}_{13}$ & $ (\Vec{\sigma}_1 + \Vec{\sigma}_2) \cdot \Vec{q_3} \; \; \Vec{q}  \cdot \Vec{q_3} \; (\pmb{\tau}_1 - \pmb{\tau}_2) $ \\

\phantom{xx}${\fett O}_{5}$  & $ q^2 \; (\Vec{\sigma}_1 \times \Vec{\sigma}_2) \cdot \Vec{q_3}  \; (\pmb{\tau}_1 \times \pmb{\tau}_2) $ & ${\fett O}_{14}$ & $ (\Vec{\sigma}_1 - \Vec{\sigma}_2) \cdot \Vec{q_3} \; \; \Vec{q}  \cdot \Vec{q_3} \; (\pmb{\tau}_1 + \pmb{\tau}_2) $ \\

\phantom{xx}${\fett O}_{6}$  & $ (\Vec{\sigma}_1 \times \Vec{\sigma}_2) \cdot \Vec{q} \;\; \Vec{q}  \cdot \Vec{q_3} \; (\pmb{\tau}_1 \times \pmb{\tau}_2) $ & ${\fett O}_{15}$ & $ (\Vec{\sigma}_1 + \Vec{\sigma}_2) \cdot \Vec{q} \; \; q_3^2 \; (\pmb{\tau}_1 - \pmb{\tau}_2) $ \\

\phantom{xx}${\fett O}_7$  & $ q^2 \; (\Vec{\sigma}_1 - \Vec{\sigma}_2) \cdot \Vec{q_3} \; (\pmb{\tau}_1 - \pmb{\tau}_2) $ & ${\fett O}_{16}$ & $ (\Vec{\sigma}_1 - \Vec{\sigma}_2) \cdot \Vec{q} \; \; q_3^2 \; (\pmb{\tau}_1 + \pmb{\tau}_2) $ \\

\phantom{xx}${\fett O}_8$  & $ q^2 \; (\Vec{\sigma}_1 + \Vec{\sigma}_2) \cdot \Vec{q_3} \; (\pmb{\tau}_1 + \pmb{\tau}_2) $ & ${\fett O}_{17}$ & $ (\Vec{\sigma}_1 - \Vec{\sigma}_2) \cdot \Vec{q_3} \;\; q_3^2 \; (\pmb{\tau}_1 - \pmb{\tau}_2) $ \\

\phantom{xx}${\fett O}_9$  & $ (\Vec{\sigma}_1 - \Vec{\sigma}_2) \cdot \Vec{q} \;\; \Vec{q}  \cdot \Vec{q_3} \; (\pmb{\tau}_1 - \pmb{\tau}_2) $ &  &  
                                        \end{tabular*}
                                        \caption{A possible choice of $17$ independent $\pi$NN operators ${\fett O}_i$ at order $Q^3$. Notice that as explained in the text, the contribution of the operator  ${\fett O}_{17}$  to the N$^4$LO 3NF can be absorbed into a redefinition of the LECs $D$ and $E_i$. }
\label{Table2}                                    
  \end{ruledtabular}
   \end{table*}
   Thus, the $D$-like N$^4$LO 3NF can be finally written in the form
   \beqa
   \label{3NFDlike}
   V &=& \frac{g_A^2}{4F_\pi^2} \frac{\vec \sigma_3 \cdot \vec q_3 \, \fett \tau_3}{q_3^2 + M_\pi^2} \cdot \Big[i  F_1  (\Vec{q} \times \Vec{k} ) \cdot \Vec{q_3} \,  (\pmb{\tau}_1 +
   \pmb{\tau}_2)
   \,+\,  i  F_2     (\Vec{\sigma}_1 - \Vec{\sigma}_2) \cdot \Vec{k} \; \Vec{q}  \cdot \Vec{q_3} \, (\pmb{\tau}_1 \times \pmb{\tau}_2)
   \,+\,  i  F_3     \Vec{k}  \cdot \Vec{q} \; (\Vec{\sigma}_1 - \Vec{\sigma}_2) \cdot \Vec{q_3} \,  (\pmb{\tau}_1 \times \pmb{\tau}_2) \nn
   && {}
   \,+\,  i  F_4     (\Vec{\sigma}_1 - \Vec{\sigma}_2) \cdot \Vec{q} \; \Vec{k}  \cdot \Vec{q_3} \; (\pmb{\tau}_1 \times \pmb{\tau}_2)
   \,+\,  F_5  q^2 \; (\Vec{\sigma}_1 \times \Vec{\sigma}_2) \cdot \Vec{q_3}  \; (\pmb{\tau}_1 \times \pmb{\tau}_2)
   \,+\,  F_6 (\Vec{\sigma}_1 \times \Vec{\sigma}_2) \cdot \Vec{q} \; \Vec{q}  \cdot \Vec{q_3} \; (\pmb{\tau}_1 \times \pmb{\tau}_2)\nn
   && {}
   \,+\,  F_7   q^2 \; (\Vec{\sigma}_1 - \Vec{\sigma}_2) \cdot \Vec{q_3} \; (\pmb{\tau}_1 - \pmb{\tau}_2)
   \,+\,  F_8   q^2 \; (\Vec{\sigma}_1 + \Vec{\sigma}_2) \cdot \Vec{q_3} \; (\pmb{\tau}_1 + \pmb{\tau}_2)
   \,+\,  F_9    (\Vec{\sigma}_1 - \Vec{\sigma}_2) \cdot \Vec{q} \; \Vec{q}  \cdot \Vec{q_3} \; (\pmb{\tau}_1 - \pmb{\tau}_2) \nn
   && {}
   \,+\,  F_{10} (\Vec{\sigma}_1 + \Vec{\sigma}_2) \cdot \Vec{q} \; \Vec{q}  \cdot \Vec{q_3} \; (\pmb{\tau}_1 + \pmb{\tau}_2)
  \,+\,  i F_{11} \Vec{\sigma}_2 \cdot \Vec{q} \; (\Vec{\sigma}_1 \times \Vec{k}) \cdot \Vec{q_3}  \; (\pmb{\tau}_1 + \pmb{\tau}_2) 
  \,+\,  i F_{12} \Vec{\sigma}_1 \cdot \Vec{q_3} \; ( \Vec{q} \times  \Vec{\sigma}_2) \cdot \Vec{q_3}  \; (\pmb{\tau}_1 \times \pmb{\tau}_2) \nn
   && {}
   \,+\,  F_{13} (\Vec{\sigma}_1 + \Vec{\sigma}_2) \cdot \Vec{q_3} \;  \Vec{q}  \cdot \Vec{q_3} \; (\pmb{\tau}_1 - \pmb{\tau}_2)
   \,+\,  F_{14}  (\Vec{\sigma}_1 - \Vec{\sigma}_2) \cdot \Vec{q_3} \;  \Vec{q}  \cdot \Vec{q_3} \; (\pmb{\tau}_1 + \pmb{\tau}_2)
   \,+\,  F_{15}  (\Vec{\sigma}_1 + \Vec{\sigma}_2) \cdot \Vec{q} \;  q_3^2 \; (\pmb{\tau}_1 - \pmb{\tau}_2) \nn
   && {}
   \,+\,  F_{16}(\Vec{\sigma}_1 - \Vec{\sigma}_2) \cdot \Vec{q} \;  q_3^2 \; (\pmb{\tau}_1 + \pmb{\tau}_2) \Big] \;+\; \mbox{5 permutations}\,,
  \eeqa
where $F_i$ denote the corresponding LECs, while $F_\pi$ refers to the pion decay constant. 
                                          
\subsection{Method II: Construction of the $\pi$NN heavy-baryon Lagrangian using reparametrization invariance}
Instead of parametrizing the 3N potential, one can proceed by constructing the non-relativistic $\pi$NN Lagrangian at order $Q^3$. Requiring $\mathcal{P}$-invariance and taking into account partial integration identities, there are $7$ possible space structures that can appear in the Lagrangian, 
\begin{equation}
\begin{split}
    (N^\dagger \vecev{\nabla} N)(N^\dagger \vecev{\nabla} N) \nabla \pi &\eqcolon X_A^+ , \quad
    \nabla (N^\dagger  N)\nabla (N^\dagger  N) \nabla \pi \eqcolon X_B^+ , \quad
    i \nabla(N^\dagger  N)(N^\dagger \vecev{\nabla} N) \nabla \pi \eqcolon X_C^- ,\\
    (N^\dagger \vecev{\nabla} \vecev{\nabla} N)(N^\dagger  N) \nabla \pi &\eqcolon X_D^+, \quad
   i (N^\dagger \vecev{\nabla} N)(N^\dagger  N) \nabla \nabla \pi \eqcolon X_E^- , \quad
    \nabla (N^\dagger  N)(N^\dagger  N) \nabla \nabla \pi \eqcolon X_F^+ ,\\
    (N^\dagger  N)(N^\dagger N) \nabla \nabla \nabla \pi &\eqcolon X_G^+ ,
\end{split}
\end{equation}
where $N^\dagger \vecev{\nabla} N \equiv N^\dagger (\vec \nabla N) - (\vec \nabla N^\dagger ) N$ and 
we have suppressed the vector indices in isospin and usual spaces. 
The factors of $i$ are included to ensure that all $7$ structures are hermitian. The superscripts $(+,-)$ show the sign the structures acquire after performing time reversal operation $\mathcal{T}$. Furthermore, the Goldstone boson nature of pions requires that all chiral-symmetry-invariant vertices in the $\pi$NN  Lagrangian involve at least one derivative acting on $\fett \pi$. Notice that we do not consider structures with time derivatives acting on $\fett \pi$, since the resulting tree-level 3NFs are suppressed by a factor of $p/m$ and thus do not contribute at N$^4$LO. Isospin symmetry restricts the isospin dependence of the considered vertices to just three structures
\begin{equation}
    1 \otimes \tau_a \eqcolon T_A^-, \quad  \tau_a \otimes 1 \eqcolon T_B^-, \quad \epsilon_{abc} \tau_b \otimes \tau_c \eqcolon T_C^+ ,
  \end{equation}
where we assume that the remaining summation index $a$ gets contracted with $\pi_a$ and the superscripts again denote the transformation property under time reversal. If the combination of a space and isospin structure $X \times T$ gives a plus (minus) sign under time-reversal, we have to include $0$ or $2$ ($1$) Pauli matrices $\Vec{\sigma}$, so that the resulting structure is invariant under $\mathcal{T}$. The spin-space indices of the structures get contracted with the Kronecker and Levi-Civita tensors. After taking into account Schouten identities, we found $66$ allowed nonrelativistic structures $O^{\RN{2}}_{i}$ listed in Table \ref{Table3}.

\begin{table*}
				\begin{ruledtabular}
					\begin{tabular*}{\textwidth}{@{\extracolsep{\fill}}llll}
\phantom{xx}$O^{\RN{2}}_{1}$\phantom{xx}& $ (N^\dagger \vecev{\nabla}_i \tau_a N)(N^\dagger \sigma_j \vecev{\nabla}_j  N)  \nabla_i  \pi_a$  \phantom{xxxxxxxxxxxxxxxxxxxxxxxx} & $O^{\RN{2}}_{34}$\phantom{xx} & $ i \epsilon_{abc} \nabla_i (N^\dagger \sigma_j \tau_b  N)(N^\dagger  \vecev{\nabla}_j \tau_c  N)  \nabla_i  \pi_a$ \\

\phantom{xx}$O^{\RN{2}}_{2}$ & $ (N^\dagger \vecev{\nabla}_i \tau_a N)(N^\dagger \sigma_j \vecev{\nabla}_i  N)  \nabla_j  \pi_a$   & $O^{\RN{2}}_{35}$ & $ i \epsilon_{abc} \nabla_i (N^\dagger  \tau_b  N)(N^\dagger \sigma_j \vecev{\nabla}_i \tau_c  N)  \nabla_j  \pi_a$ \\

\phantom{xx}$O^{\RN{2}}_{3}$ & $ (N^\dagger \vecev{\nabla}_i \tau_a N)(N^\dagger \sigma_i \vecev{\nabla}_j  N)  \nabla_j  \pi_a$    & $O^{\RN{2}}_{36}$ & $ i \epsilon_{abc} \nabla_i (N^\dagger  \tau_b  N)(N^\dagger \sigma_i \vecev{\nabla}_j \tau_c  N)  \nabla_j  \pi_a$ \\

\phantom{xx}$O^{\RN{2}}_{4}$ & $ (N^\dagger \vecev{\nabla}_i N)(N^\dagger \sigma_j \vecev{\nabla}_i \tau_a N)  \nabla_j  \pi_a$   & $O^{\RN{2}}_{37}$ & $ i \epsilon_{abc} \nabla_i (N^\dagger  \tau_b  N)(N^\dagger \sigma_j \vecev{\nabla}_j \tau_c  N)  \nabla_i  \pi_a$ \\

\phantom{xx}$O^{\RN{2}}_{5}$ & $ (N^\dagger \vecev{\nabla}_i  N)(N^\dagger \sigma_i \vecev{\nabla}_j \tau_a N)  \nabla_j  \pi_a$ & $O^{\RN{2}}_{38}$ & $ (N^\dagger \sigma_j \vecev{\nabla}_i \vecev{\nabla}_i N)(N^\dagger \tau_a  N)  \nabla_j  \pi_a$\\

\phantom{xx}$O^{\RN{2}}_{6}$ & $ (N^\dagger \vecev{\nabla}_i  N)(N^\dagger \sigma_j \vecev{\nabla}_j \tau_a  N)  \nabla_i  \pi_a$  & $O^{\RN{2}}_{39}$ & $ (N^\dagger \sigma_i \vecev{\nabla}_i \vecev{\nabla}_j N)(N^\dagger \tau_a  N)  \nabla_j  \pi_a$\\

\phantom{xx}$O^{\RN{2}}_{7}$ & $ \epsilon_{ijk} \epsilon_{abc} (N^\dagger \vecev{\nabla}_i \tau_b N)(N^\dagger \vecev{\nabla}_j \tau_c N)  \nabla_k  \pi_a$   & $O^{\RN{2}}_{40}$ & $ (N^\dagger \vecev{\nabla}_i \vecev{\nabla}_i N)(N^\dagger \sigma_j \tau_a  N)  \nabla_j  \pi_a$\\

\phantom{xx}$O^{\RN{2}}_{8}$ & $ \epsilon_{ijk} \epsilon_{abc} (N^\dagger \sigma_j \vecev{\nabla}_i \tau_b N) (N^\dagger \sigma_k \vecev{\nabla}_l \tau_c N)  \nabla_l  \pi_a$ & $O^{\RN{2}}_{41}$ & $ (N^\dagger \vecev{\nabla}_i \vecev{\nabla}_j N)(N^\dagger \sigma_i \tau_a  N)  \nabla_j  \pi_a$\\

\phantom{xx}$O^{\RN{2}}_{9}$ & $ \epsilon_{ijk} \epsilon_{abc} (N^\dagger \sigma_l \vecev{\nabla}_i \tau_b N) (N^\dagger \sigma_l \vecev{\nabla}_j \tau_c N)  \nabla_k  \pi_a$ & $O^{\RN{2}}_{42}$ & $ (N^\dagger \sigma_j \vecev{\nabla}_i \vecev{\nabla}_i \tau_a N)(N^\dagger   N)  \nabla_j  \pi_a$\\

\phantom{xx}$O^{\RN{2}}_{10}$ & $ \epsilon_{ijk} \epsilon_{abc} (N^\dagger \sigma_l \vecev{\nabla}_i \tau_b N) (N^\dagger \sigma_k \vecev{\nabla}_j \tau_c N)  \nabla_l  \pi_a$ & $O^{\RN{2}}_{43}$ & $ (N^\dagger \sigma_i \vecev{\nabla}_i \vecev{\nabla}_j \tau_a N)(N^\dagger   N)  \nabla_j  \pi_a$\\

\phantom{xx}$O^{\RN{2}}_{11}$ & $ \epsilon_{ijk} \epsilon_{abc} (N^\dagger \sigma_l \vecev{\nabla}_l \tau_b N) (N^\dagger \sigma_j \vecev{\nabla}_i \tau_c N)  \nabla_k  \pi_a$ & $O^{\RN{2}}_{44}$ & $ (N^\dagger \vecev{\nabla}_i \vecev{\nabla}_i \tau_a N)(N^\dagger \sigma_j   N)  \nabla_j  \pi_a$\\

\phantom{xx}$O^{\RN{2}}_{12}$ & $\nabla_i (N^\dagger  \tau_a N) \nabla_i (N^\dagger \sigma_j N)  \nabla_j  \pi_a$    & $O^{\RN{2}}_{45}$ & $ (N^\dagger \vecev{\nabla}_i \vecev{\nabla}_j \tau_a N)(N^\dagger \sigma_i   N)  \nabla_j  \pi_a$\\

\phantom{xx}$O^{\RN{2}}_{13}$ & $\nabla_i (N^\dagger  \tau_a N) \nabla_j (N^\dagger \sigma_i N)  \nabla_j  \pi_a$  & $O^{\RN{2}}_{46}$ & $ - \epsilon_{ijk} \epsilon_{abc} (N^\dagger \sigma_l \vecev{\nabla}_l \vecev{\nabla}_i \tau_b N)(N^\dagger \sigma_j \tau_c  N)  \nabla_k  \pi_a$\\

\phantom{xx}$O^{\RN{2}}_{14}$ & $\nabla_i (N^\dagger   N) \nabla_i (N^\dagger \sigma_j \tau_a N)  \nabla_j  \pi_a$ & $O^{\RN{2}}_{47}$ & $ - \epsilon_{ijk} \epsilon_{abc} (N^\dagger \sigma_j \vecev{\nabla}_i \vecev{\nabla}_l \tau_b N)(N^\dagger \sigma_l \tau_c  N)  \nabla_k  \pi_a$\\

\phantom{xx}$O^{\RN{2}}_{15}$ & $\nabla_i (N^\dagger  N) \nabla_j (N^\dagger \sigma_j  \tau_a N)  \nabla_i  \pi_a$ & $O^{\RN{2}}_{48}$ & $ \epsilon_{ijk} \epsilon_{abc} (N^\dagger \sigma_j \vecev{\nabla}_i \vecev{\nabla}_l \tau_b N)(N^\dagger \sigma_k \tau_c  N)  \nabla_l  \pi_a$\\

\phantom{xx}$O^{\RN{2}}_{16}$ & $ \epsilon_{ijk} \epsilon_{abc}  \nabla_i (N^\dagger \sigma_j \tau_b  N) \nabla_l (N^\dagger \sigma_k  \tau_c N)  \nabla_l  \pi_a$  & $O^{\RN{2}}_{49}$ & $ - i \epsilon_{ijk} (N^\dagger \sigma_l \vecev{\nabla}_i  N)(N^\dagger \sigma_j \tau_a N)  \nabla_k  \nabla_l  \pi_a$\\

\phantom{xx}$O^{\RN{2}}_{17}$ & $ \epsilon_{ijk} \epsilon_{abc}  \nabla_l (N^\dagger \sigma_l \tau_b  N) \nabla_i (N^\dagger \sigma_j  \tau_c N)  \nabla_k  \pi_a$ & $O^{\RN{2}}_{50}$ & $ - i \epsilon_{ijk} (N^\dagger \sigma_j \vecev{\nabla}_i  N)(N^\dagger \sigma_l \tau_a N)  \nabla_k  \nabla_l  \pi_a$\\

\phantom{xx}$O^{\RN{2}}_{18}$ & $ i \epsilon_{ijk} \nabla_i (N^\dagger   N)(N^\dagger \vecev{\nabla}_j \tau_a N)  \nabla_k  \pi_a$  & $O^{\RN{2}}_{51}$ & $ i \epsilon_{ijk} (N^\dagger \sigma_j \vecev{\nabla}_i  N)(N^\dagger \sigma_k \tau_a N)  \nabla^2   \pi_a$\\

\phantom{xx}$O^{\RN{2}}_{19}$ & $ - i \epsilon_{ijk} \nabla_i (N^\dagger \sigma_l  N)(N^\dagger \sigma_j \vecev{\nabla}_l \tau_a  N)  \nabla_k  \pi_a$ & $O^{\RN{2}}_{52}$ & $ - i \epsilon_{ijk} (N^\dagger \sigma_l \vecev{\nabla}_i  \tau_a N)(N^\dagger \sigma_j  N)  \nabla_k  \nabla_l  \pi_a$\\

\phantom{xx}$O^{\RN{2}}_{20}$ & $ i \epsilon_{ijk} \nabla_i (N^\dagger \sigma_l  N)(N^\dagger \sigma_l \vecev{\nabla}_j \tau_a  N)  \nabla_k  \pi_a$ & $O^{\RN{2}}_{53}$ & $ - i \epsilon_{ijk} (N^\dagger \sigma_j \vecev{\nabla}_i \tau_a N)(N^\dagger \sigma_l  N)  \nabla_k  \nabla_l  \pi_a$\\

\phantom{xx}$O^{\RN{2}}_{21}$ & $ i \epsilon_{ijk} \nabla_i (N^\dagger \sigma_l  N)(N^\dagger \sigma_k \vecev{\nabla}_j \tau_a  N)  \nabla_l  \pi_a$  & $O^{\RN{2}}_{54}$ & $ i \epsilon_{ijk} (N^\dagger \sigma_j \vecev{\nabla}_i \tau_a N)(N^\dagger \sigma_k  N)  \nabla^2   \pi_a$\\

\phantom{xx}$O^{\RN{2}}_{22}$ & $ - i \epsilon_{ijk} \nabla_i (N^\dagger \sigma_j  N)(N^\dagger \sigma_l \vecev{\nabla}_l \tau_a  N)  \nabla_k  \pi_a$  & $O^{\RN{2}}_{55}$ & $ i \epsilon_{abc} (N^\dagger \sigma_i \vecev{\nabla}_i \tau_b N)(N^\dagger \tau_c  N)  \nabla^2   \pi_a$\\

\phantom{xx}$O^{\RN{2}}_{23}$ & $ i \epsilon_{ijk} \nabla_i (N^\dagger \sigma_j  N)(N^\dagger \sigma_k \vecev{\nabla}_l \tau_a  N)  \nabla_l  \pi_a$  & $O^{\RN{2}}_{56}$ & $ i \epsilon_{abc} (N^\dagger \sigma_j \vecev{\nabla}_i \tau_b N)(N^\dagger \tau_c  N)  \nabla_i \nabla_j   \pi_a$\\

\phantom{xx}$O^{\RN{2}}_{24}$ & $ - i \epsilon_{ijk} \nabla_i (N^\dagger \sigma_j  N)(N^\dagger \sigma_l \vecev{\nabla}_k \tau_a  N)  \nabla_l  \pi_a$  & $O^{\RN{2}}_{57}$ & $ i \epsilon_{abc} (N^\dagger \vecev{\nabla}_i \tau_b N)(N^\dagger \sigma_i \tau_c  N)   \nabla^2   \pi_a$\\

\phantom{xx}$O^{\RN{2}}_{25}$ & $ i \epsilon_{ijk} \nabla_i (N^\dagger \tau_a N)(N^\dagger \vecev{\nabla}_j   N)  \nabla_k  \pi_a$  & $O^{\RN{2}}_{58}$ & $ i \epsilon_{abc} (N^\dagger \vecev{\nabla}_i \tau_b N)(N^\dagger \sigma_j \tau_c  N)   \nabla_i \nabla_j   \pi_a$\\

\phantom{xx}$O^{\RN{2}}_{26}$ & $ - i \epsilon_{ijk} \nabla_i (N^\dagger \sigma_l \tau_a  N)(N^\dagger \sigma_j \vecev{\nabla}_l   N)  \nabla_k  \pi_a$  & $O^{\RN{2}}_{59}$ & $ \nabla_i (N^\dagger \sigma_i \tau_a  N)  (N^\dagger N)  \nabla^2 \pi_a$ \\

\phantom{xx}$O^{\RN{2}}_{27}$ & $ i \epsilon_{ijk} \nabla_i (N^\dagger \sigma_l \tau_a N)(N^\dagger \sigma_l \vecev{\nabla}_j   N)  \nabla_k  \pi_a$  & $O^{\RN{2}}_{60}$ & $ \nabla_i (N^\dagger \sigma_j  \tau_a N)  (N^\dagger N)  \nabla_i \nabla_j \pi_a$ \\

\phantom{xx}$O^{\RN{2}}_{28}$ & $ i \epsilon_{ijk} \nabla_i (N^\dagger \sigma_l \tau_a  N)(N^\dagger \sigma_k \vecev{\nabla}_j   N)  \nabla_l  \pi_a$  & $O^{\RN{2}}_{61}$ & $ \nabla_i (N^\dagger   \tau_a N)  (N^\dagger \sigma_i N)  \nabla^2 \pi_a$ \\

\phantom{xx}$O^{\RN{2}}_{29}$ & $ - i \epsilon_{ijk} \nabla_i (N^\dagger \sigma_j \tau_a N)(N^\dagger \sigma_l \vecev{\nabla}_l   N)  \nabla_k  \pi_a$  & $O^{\RN{2}}_{62}$ & $ \nabla_i (N^\dagger  \tau_a  N)  (N^\dagger \sigma_j  N)  \nabla_i \nabla_j \pi_a$ \\

\phantom{xx}$O^{\RN{2}}_{30}$ & $ i \epsilon_{ijk} \nabla_i (N^\dagger \sigma_j \tau_a  N)(N^\dagger \sigma_k \vecev{\nabla}_l   N)  \nabla_l  \pi_a$  & $O^{\RN{2}}_{63}$ & $ - \epsilon_{ijk} \epsilon_{abc} \nabla_i (N^\dagger \sigma_j \tau_b  N)  (N^\dagger \sigma_l \tau_c  N)  \nabla_k \nabla_l \pi_a$ \\

\phantom{xx}$O^{\RN{2}}_{31}$ & $ - i \epsilon_{ijk} \nabla_i (N^\dagger \sigma_j \tau_a  N)(N^\dagger \sigma_l \vecev{\nabla}_k   N)  \nabla_l  \pi_a$  & $O^{\RN{2}}_{64}$ & $  (N^\dagger \sigma_i  N)  (N^\dagger \tau_a N) \nabla_i \nabla^2  \pi_a$ \\

\phantom{xx}$O^{\RN{2}}_{32}$ & $ i \epsilon_{abc} \nabla_i (N^\dagger \sigma_i \tau_b  N)(N^\dagger  \vecev{\nabla}_j \tau_c  N)  \nabla_j  \pi_a$  & $O^{\RN{2}}_{65}$ & $  (N^\dagger \sigma_i \tau_a N)  (N^\dagger  N) \nabla_i \nabla^2  \pi_a$ \\

\phantom{xx}$O^{\RN{2}}_{33}$ & $ i \epsilon_{abc} \nabla_i (N^\dagger \sigma_j \tau_b  N)(N^\dagger  \vecev{\nabla}_i \tau_c  N)  \nabla_j  \pi_a$ & $O^{\RN{2}}_{66}$ & $ \epsilon_{ijk} \epsilon_{abc} (N^\dagger \sigma_i \tau_b N)  (N^\dagger \sigma_j \tau_c N) \nabla_k \nabla^2  \pi_a$
                                       \end{tabular*}
                                        \caption{$66$ possible structures in the non-relativistic $\pi$NN Lagrangian at order $Q^3$.}
\label{Table3}                                    
  \end{ruledtabular}
   \end{table*}

The operators listed in Table~\ref{Table3} are not independent of each other since we have not yet taken into account Fierz identities (or, equivalently, the implications of the Pauli principle): 
\begin{equation}
    (\sigma_\mu \otimes \sigma_\nu)(\tau_\alpha \otimes \tau_\beta) = - \sum_{\rho,\sigma} \sum_{\gamma,\delta} (\sigma_\rho \otimes \sigma_\sigma)(\tau_\gamma \otimes \tau_\delta) \frac{\mathrm{Tr}(\sigma_\mu \sigma_\sigma \sigma_\nu \sigma_\rho)}{4} \frac{\mathrm{Tr}(\tau_\alpha \tau_\delta \tau_\beta \tau_\gamma)}{4},
\end{equation}
where $\sigma_0$ and  $\tau_0$ denote the identity matrices in the spin and isospin spaces and the derivatives acting on the nucleon fields have to be rearranged according to $\vecev{\nabla}^a \rightarrow \frac{1}{2}(\vec \nabla^b + \vecev{\nabla}^b - \vec \nabla^a + \vecev{\nabla}^a)$ and $\nabla^a \rightarrow \frac{1}{2}(\vec \nabla^b + \vecev{\nabla}^b + \vec \nabla^a - \vecev{\nabla}^a)$ with $b \neq a$, where $a,b\in \{1,2\}$ represents the nucleon bilinear the derivative is acting on.
After taking into account $66$ Fierz identities, we are still left with  $31$ operators $O^{\RN{2}}_{i}$\footnote{We have verified these results by constructing and subsequently antisymmetrizing the potentials corresponding to the $31$ and $66$ structures and comparing the resulting expressions with each other.}. As before, the choice of the independent operators is not unique.
One possible choice of independent operators corresponds to keeping only isospin structures $T_A^-$ and $T_B^-$, e.g.:
\begin{equation}
\begin{split}
    &O^{\RN{2}}_{4}, 
    \quad O^{\RN{2}}_{5}, 
    \quad O^{\RN{2}}_{6}, 
    \quad  O^{\RN{2}}_{14},
    \quad  O^{\RN{2}}_{15},
    \quad  O^{\RN{2}}_{19},
    \quad  O^{\RN{2}}_{20},
    \quad  O^{\RN{2}}_{22},
    \quad  O^{\RN{2}}_{23},
    \quad  O^{\RN{2}}_{25},
    \quad  O^{\RN{2}}_{26},
    \quad  O^{\RN{2}}_{27},
    \quad O^{\RN{2}}_{28},
    \quad O^{\RN{2}}_{29},
    \quad O^{\RN{2}}_{30},
    \quad O^{\RN{2}}_{31}, \\
    & O^{\RN{2}}_{38},
    \quad O^{\RN{2}}_{39},
    \quad O^{\RN{2}}_{40},
    \quad O^{\RN{2}}_{41},
    \quad O^{\RN{2}}_{42},
    \quad O^{\RN{2}}_{43},
    \quad O^{\RN{2}}_{44},
    \quad O^{\RN{2}}_{45}, 
    \quad O^{\RN{2}}_{51},
    \quad O^{\RN{2}}_{52},
    \quad O^{\RN{2}}_{53},
    \quad O^{\RN{2}}_{54},
    \quad O^{\RN{2}}_{61},
    \quad O^{\RN{2}}_{62},
    \quad O^{\RN{2}}_{65}.
\end{split}
\label{eq:5}
\end{equation}

The reason for ending up with $31$ instead of $17$ operators is that we have not yet taken into account constraints from Poincaré invariance. To account for these constraints on the level of the effective Lagrangian, we require the constructed Lagrangian to be reparametrization invariant \cite{Luke:1992cs}, see Ref.~\cite{Epelbaum:2000kv} for the application to the NN Lagrangian at order $Q^2$.\footnote{Clearly, reparametrization invariance of the order-$Q^3$ $\pi$NN Lagrangian just amounts to constraining the interactions such that the resulting 3NF is independent of the total momentum of the nucleons.} The reparametrization-invariant Lagrangian is given by $17$ linear combinations of the operators $O^{\RN{2}}_i$. The Feynman rules for these $17$ linear combinations, ${\rm FR}[ \ldots ]$, can be expressed in terms of the operators $O_i^a$ from Table~\ref{Table2}, where $a$ denotes the pion isospin index, as follows:  
\beqa
\label{FR}
{\rm FR}\big[ 8 O^{\RN{2}}_{14} \big] &=& 4 O_7^a + 4O_8^a - O_{17}^a , \nn
{\rm FR}\big[ 8 O^{\RN{2}}_{15}\big] &=&4 O_9^a + 4O_{10}^a - 2O_{13}^a - 2O_{14}^a + 2O_{15}^a +2O_{16}^a - O_{17}^a , \nn
{\rm FR}\big[ O^{\RN{2}}_{20} + O^{\RN{2}}_{27}\big] &=&  -4 O_1^a  , \nn
{\rm FR}\big[ O^{\RN{2}}_{19} + O^{\RN{2}}_{29}\big] &=&  -2 O_2^a +2 O_3^a +2 O_{11}^a   , \nn
{\rm FR}\big[2 O^{\RN{2}}_{4} - O^{\RN{2}}_{40} - O^{\RN{2}}_{42} \big] &=&  2 O_5^a -2 O_8^a   , \nn
{\rm FR}\big[  O^{\RN{2}}_{5} + O^{\RN{2}}_{6} - O^{\RN{2}}_{41}  - O^{\RN{2}}_{43}\big] &=& 2 O_6^a -2 O_{10}^a    , \nn
{\rm FR}\big[  O^{\RN{2}}_{38} + O^{\RN{2}}_{40} + O^{\RN{2}}_{42}  + O^{\RN{2}}_{44}\big] &=&  2 O_8^a    , \nn
{\rm FR}\big[ O^{\RN{2}}_{39} + O^{\RN{2}}_{41} + O^{\RN{2}}_{43}  + O^{\RN{2}}_{45}\big] &=&  2 O_{10}^a    , \nn
{\rm FR}\big[ O^{\RN{2}}_{52} + O^{\RN{2}}_{53}\big] &=&   O_{12}^a   , \nn
{\rm FR}\big[ O^{\RN{2}}_{23} - O^{\RN{2}}_{30} - 2 O^{\RN{2}}_{53}  - O^{\RN{2}}_{54}\big] &=&  -4 O_{2}^a - O_{12}^a - O_{14}^a  , \nn
{\rm FR}\big[ 2O^{\RN{2}}_{28} - 4 O^{\RN{2}}_{30} - 2 O^{\RN{2}}_{31}  + 2O^{\RN{2}}_{41} + 2O^{\RN{2}}_{45} - 2O^{\RN{2}}_{53} - 2O^{\RN{2}}_{54}\big] &=&  -4 O_{2}^a -4 O_{4}^a + 2O_{10}^a  - O_{12}^a -2 O_{14}^a -2 O_{16}^a     , \nn
{\rm FR}\big[ 2O^{\RN{2}}_{26} - 2 O^{\RN{2}}_{29} - 2 O^{\RN{2}}_{30}  + O^{\RN{2}}_{40} + O^{\RN{2}}_{44} - 2 O^{\RN{2}}_{53} - O^{\RN{2}}_{54}\big] &=&  -4 O_{3}^a + O_{8}^a   - O_{12}^a -2 O_{14}^a    , \nn
{\rm FR}\big[  2O^{\RN{2}}_{22} + 2 O^{\RN{2}}_{29} +2  O^{\RN{2}}_{30}  - O^{\RN{2}}_{40} - O^{\RN{2}}_{44} + 2 O^{\RN{2}}_{53} + O^{\RN{2}}_{54}\big] &=& 4 O_{2}^a - O_{8}^a   +  4 O_{11}^a + O_{12}^a + 2 O_{14}^a    , \nn
{\rm FR}\big[ O^{\RN{2}}_{51} + O^{\RN{2}}_{54}\big] &=& 2 O_{16}^a   , \nn
{\rm FR}\big[  4 O^{\RN{2}}_{61}\big] &=&  -2 O_{15}^a + 2O_{16}^a - O_{17}^a   , \nn
{\rm FR}\big[  4 O^{\RN{2}}_{62}\big] &=&  -2 O_{13}^a + 2O_{14}^a - O_{17}^a   , \nn
{\rm FR}\big[ 2 O^{\RN{2}}_{65}\big] &=&     - O_{17}^a . 
\eeqa
Thus, both approaches lead to the same expressions for the 3NF.

\begin{table*}
				\begin{ruledtabular}
					\begin{tabular*}{\textwidth}{@{\extracolsep{\fill}}llclllll}
\phantom{xxxxxx}$ I_4 \otimes \gamma_5 \gamma$\phantom{xxx} & $O^{\RN{3}}_1$  \phantom{x} & $(\Bar{\psi} \gamma_5 \gamma^\mu \tau_a \psi) \partial^2 (\Bar{\psi} \psi) (\partial_\mu \pi_a)$ \\ 
  & $O^{\RN{3}}_2$ & $(\Bar{\psi} \gamma_5 \gamma^\mu \psi) \partial^2 (\Bar{\psi} \tau_a \psi) (\partial_\mu \pi_a)$ \\ 
   & $O^{\RN{3}}_3$ & $(\Bar{\psi} \tau_a \psi) \partial^2 (\Bar{\psi} \gamma_5 \gamma^\mu \psi) (\partial_\mu \pi_a)$ \\ 
 & $O^{\RN{3}}_4$ & $(\Bar{\psi}  \psi) \partial^2 (\Bar{\psi} \gamma_5 \gamma^\mu \tau_a \psi) (\partial_\mu \pi_a)$ \\ 
   & $O^{\RN{3}}_5$ & $(\Bar{\psi} \tau_a \vecev{\partial}^2 \psi)(\Bar{\psi} \gamma_5 \gamma^\mu  \psi) (\partial_\mu \pi_a)- (\Bar{\psi} \tau_a \vecev{\partial}_\alpha \psi)(\Bar{\psi} \gamma_5 \gamma^\mu   \vecev{\partial}^\alpha \psi) (\partial_\mu \pi_a)$ \\ 
   & $O^{\RN{3}}_6$ & $(\Bar{\psi} \vecev{\partial}^2 \psi)(\Bar{\psi} \gamma_5 \gamma^\mu \tau_a  \psi) (\partial_\mu \pi_a)- (\Bar{\psi} \vecev{\partial}_\alpha \psi)(\Bar{\psi} \gamma_5 \gamma^\mu \tau_a  \vecev{\partial}^\alpha \psi) (\partial_\mu \pi_a)$ \\ 
& $O^{\RN{3}}_7$ & $(\Bar{\psi} \tau_a \vecev{\partial}_\mu  \psi)(\Bar{\psi} \gamma_5 \gamma^\mu \vecev{\partial}^\alpha \psi) (\partial_\alpha \pi_a) - (\Bar{\psi} \tau_a \vecev{\partial}_\mu \vecev{\partial}^\alpha  \psi)(\Bar{\psi} \gamma_5 \gamma^\mu \psi) (\partial_\alpha \pi_a)$ \\ 
& $O^{\RN{3}}_8$ & $(\Bar{\psi} \vecev{\partial}_\mu  \psi)(\Bar{\psi} \gamma_5 \gamma^\mu \tau_a  \vecev{\partial}^\alpha \psi) (\partial_\alpha \pi_a) - (\Bar{\psi} \vecev{\partial}_\mu \vecev{\partial}^\alpha  \psi)(\Bar{\psi} \gamma_5 \gamma^\mu \tau_a  \psi) (\partial_\alpha \pi_a)$\phantom{xxxxxx} \\ 
 & $O^{\RN{3}}_{11}$ & $(\Bar{\psi}  \gamma_5 \gamma^\mu \tau_a \psi) \partial^\alpha (\Bar{\psi} \psi) (\partial_\mu \partial_\alpha \pi_a)$ \\ 
  & $O^{\RN{3}}_{12}$ & $(\Bar{\psi}  \gamma_5 \gamma^\mu \psi) \partial^\alpha (\Bar{\psi} \tau_a  \psi) (\partial_\mu \partial_\alpha \pi_a)$ \\    
\midrule
\phantom{xxxxxx}$ \gamma_5 \gamma \otimes \sigma $ & $O^{\RN{3}}_{13}$ & $i (\Bar{\psi} \sigma^{\mu \nu} \tau_a \psi) \partial_\nu (\Bar{\psi}   \gamma_5 \gamma^\alpha \vecev{\partial}_\mu \psi) ( \partial_\alpha  \pi_a) $ \\
& $O^{\RN{3}}_{14}$ & $i (\Bar{\psi} \sigma^{\mu \nu}  \psi) \partial_\nu (\Bar{\psi}   \gamma_5 \gamma^\alpha \tau_a \vecev{\partial}_\mu \psi) ( \partial_\alpha  \pi_a) $ \\
  & $O^{\RN{3}}_{15}$ & $i (\Bar{\psi} \sigma^{\mu \nu} \tau_a \psi) \partial_\mu (\Bar{\psi}   \gamma_5 \gamma_\nu \vecev{\partial}^\alpha \psi) ( \partial_\alpha  \pi_a) - i (\Bar{\psi} \sigma^{\mu \nu} \tau_a \vecev{\partial}^\alpha \psi) \partial_\mu (\Bar{\psi}   \gamma_5 \gamma_\nu  \psi) ( \partial_\alpha  \pi_a) $ \\
    & $O^{\RN{3}}_{16}$ & $i (\Bar{\psi} \sigma^{\mu \nu}  \psi) \partial_\mu (\Bar{\psi}   \gamma_5 \gamma_\nu \tau_a \vecev{\partial}^\alpha \psi) ( \partial_\alpha  \pi_a) - i (\Bar{\psi} \sigma^{\mu \nu} \vecev{\partial}^\alpha \psi) \partial_\mu (\Bar{\psi}   \gamma_5 \gamma_\nu \tau_a \psi) ( \partial_\alpha  \pi_a) $ \\
  & $O^{\RN{3}}_{19}$ & $i (\Bar{\psi} \sigma^{\mu \nu} \tau_a \psi) \partial^\alpha (\Bar{\psi}   \gamma_5 \gamma_\nu \vecev{\partial}_\mu \psi) ( \partial_\alpha  \pi_a) $ \\
 & $O^{\RN{3}}_{20}$ & $i (\Bar{\psi} \sigma^{\mu \nu} \psi) \partial^\alpha (\Bar{\psi}   \gamma_5 \gamma_\nu \tau_a \vecev{\partial}_\mu \psi) ( \partial_\alpha  \pi_a) $ \\
  & $O^{\RN{3}}_{21}$ & $i (\Bar{\psi} \sigma^{\mu \nu} \tau_a \vecev{\partial}_\alpha \psi) \partial_\nu (\Bar{\psi}   \gamma_5 \gamma^\alpha  \psi) ( \partial_\mu  \pi_a) $ \\
    & $O^{\RN{3}}_{22}$ & $i (\Bar{\psi} \sigma^{\mu \nu}  \vecev{\partial}_\alpha \psi) \partial_\nu (\Bar{\psi}   \gamma_5 \gamma^\alpha  \tau_a \psi) ( \partial_\mu  \pi_a) $ \\
   & $O^{\RN{3}}_{23}$ & $i  (\Bar{\psi} \gamma_5 \gamma^\alpha  \tau_a  \vecev{\partial}_\nu \psi) (\Bar{\psi} \sigma^{\mu \nu}\psi) ( \partial_\alpha  \partial_\mu  \pi_a) $ \\
& $O^{\RN{3}}_{24}$ & $i  (\Bar{\psi} \gamma_5 \gamma^\alpha \vecev{\partial}_\nu \psi) (\Bar{\psi} \sigma^{\mu \nu} \tau_a  \psi) ( \partial_\alpha  \partial_\mu  \pi_a) $ \\
& $O^{\RN{3}}_{25}$ & $i  (\Bar{\psi} \gamma_5 \gamma_\nu \tau_a \psi) (\Bar{\psi} \sigma^{\mu \nu}  \vecev{\partial}^\alpha  \psi) ( \partial_\alpha  \partial_\mu  \pi_a) - i  (\Bar{\psi} \gamma_5 \gamma_\nu \tau_a \vecev{\partial}^\alpha \psi) (\Bar{\psi} \sigma^{\mu \nu}  \psi) ( \partial_\alpha  \partial_\mu  \pi_a) $ \\
& $O^{\RN{3}}_{26}$ & $i  (\Bar{\psi} \gamma_5 \gamma_\nu \psi) (\Bar{\psi} \sigma^{\mu \nu} \tau_a  \vecev{\partial}^\alpha  \psi) ( \partial_\alpha  \partial_\mu  \pi_a) - i  (\Bar{\psi} \gamma_5 \gamma_\nu  \vecev{\partial}^\alpha \psi) (\Bar{\psi} \sigma^{\mu \nu} \tau_a  \psi) ( \partial_\alpha  \partial_\mu  \pi_a) $ \\
 & $O^{\RN{3}}_{27}$ & $i  (\Bar{\psi} \gamma_5 \gamma_\nu  \tau_a  \psi) \partial^\alpha   (\Bar{\psi} \sigma^{\mu \nu}  \vecev{\partial}_\mu \psi) ( \partial_\alpha  \pi_a) $ \\
  & $O^{\RN{3}}_{28}$ & $i  (\Bar{\psi} \gamma_5 \gamma_\nu  \psi) \partial^\alpha (\Bar{\psi} \sigma^{\mu \nu} \tau_a \vecev{\partial}_\mu \psi) ( \partial_\alpha  \pi_a) $ \\
   \midrule
\phantom{xxxxxx}$ \gamma_5 \otimes \gamma $ & $O^{\RN{3}}_{29}$ & $(\Bar{\psi}  \gamma^\mu \tau_a \psi) \partial^\alpha (\Bar{\psi} \gamma_5  \vecev{\partial}_\mu \psi)  (\partial_\alpha \pi_a)$ \\ 
    & $O^{\RN{3}}_{30}$ & $(\Bar{\psi}  \gamma^\mu \psi) \partial^\alpha (\Bar{\psi} \gamma_5 \tau_a \vecev{\partial}_\mu \psi)  (\partial_\alpha \pi_a)$ 
                                      \end{tabular*}
                                        \caption{The list of 30 covariant structures in the order-$Q^3$ $\pi$NN Lagrangian.}
\label{Table5}                                    
  \end{ruledtabular}
   \end{table*}

   \subsection{Method III: Non-relativistic reduction of the covariant Lagrangian}
\label{sec:Method3}
   
   As an independent check of the achieved results, we now follow an alternative path by starting from the covariant Lagrangian and subsequently performing its non-relativistic reduction. We begin by constructing the relevant covariant structures, which fulfill all the relevant symmetry requirements and are listed in Table~\ref{Table5}, see Refs.~\cite{Xiao:2018jot,Sun:2025zuk} for a related discussion. Notice that the structures $O^{\RN{3}}_{5,\ldots ,8}$, $O^{\RN{3}}_{15,16}$ and  $O^{\RN{3}}_{25,26}$ are given as linear combinations of two covariant structures. 
   The corresponding linear combinations with the ``$+$'' sign give rise to terms $\propto m$ after performing the nonrelativistic reduction. They thus must be multiplied with a factor of $1/m$ to ensure the existence of the proper static limit, see, e.g., Ref.~\cite{Fettes:2000gb} for similar considerations in the single-nucleon sector. In the employed counting scheme for the nucleon mass, such vertices start contributing to the 3NF at order N$^5$LO, which is beyond the accuracy of this study. We, therefore, do not list such terms in Table~\ref{Table5}. 

   After performing the non-relativistic reduction, see, e.g.,~Ref.~\cite{Girlanda:2010ya}, the leading (static) contributions of the operators $O^{\RN{3}}_i$ can be expressed in terms of $31$ nonrelativistic operators $O^{\RN{2}}_i$ specified in Eq.~\eqref{eq:5} as detailed in Appendix~\ref{App}.
   Moreover, one observes that all linear combinations of the operators $O_{i}^{\RN{2}}$ appearing on the right-hand sides of Eq.~\eqref{NRreduction} are expressible in terms of $17$ linear combinations appearing in Eq.~\eqref{FR}.
   We thus conclude that all three approaches yield consistent results for the $D$-like 3NF at N$^4$LO.

   \section{Isospin decomposition}
   \label{sec:3NFIsospin}

The LECs $F_{1, \ldots , 16}$ entering the $1\pi$-contact 3NF at N$^4$LO, see Eq.~(\ref{3NFDlike}), are unknown and will have to be determined from experimental data. Clearly, three-nucleon scattering provides the most natural and computationally accessible process to constrain 3NFs. However, in the limit of exact isospin symmetry, nucleon-deuteron (Nd) scattering is only sensitive to the isospin-$1/2$ component of the 3NF. For purely short-range 3NF, it is known that $11$ linear combinations of $13$ LECs at N$^4$LO contribute in the isospin-$1/2$ channel, while $2$ combinations appear solely in the isospin-$3/2$ channel \cite{Filandri:2026ori} and thus cannot be reliably determined from Nd scattering. It is now interesting to analyze the situation with the subleading $1\pi$-contact 3NF. 
To this aim, we use the corresponding isospin projection operators given by
\begin{equation}
    P_{1/2} = \frac{1}{2} -  \frac{{\fett \tau_1} \cdot {\fett \tau_2} + {\fett \tau_1} \cdot {\fett \tau_3} + {\fett \tau_2} \cdot {\fett \tau_3}}{6} \;\;\;\;\text{and}\;\;\;\; P_{3/2} = 1 - P_{1/2}.
\end{equation}
We then obtain the contribution of an operator $V_i$ entering the 3NF in Eq.~(\ref{3NFDlike}), $V = \sum_{i=1}^{16}F_i V_i$, to the $T=1/2$ and the $T=3/2$ channel by computing $P_{1/2} V_i P_{1/2} \equiv (V_i)_{1/2}$ and $P_{3/2} V_i P_{3/2} \equiv (V_i)_{3/2}$, respectively. For the $16$ operators appearing in the 3NF, we find 
\begin{equation}
    (V_i)_{1/2} = V_i, \;\; (V_i)_{3/2} = 0 \;\;\; \text{for} \;\; i \in \{ 2,3,4,5,6,7,9,12,13,15\} ,
\end{equation}
while the remaining $6$ operators contribute to both isospin-$1/2$ and $3/2$ channels. Contrary to the case of the purely contact interactions, we are unable to find linear combinations of these six operators, which contribute solely to the isospin-$1/2$ or $3/2$ channel. We, therefore, conclude that all $16$ LECs $F_i$ can, at least in principle, be determined from nucleon-deuteron scattering observables.  

\section{Unitary ambiguity of the two- and three-nucleon forces}
\label{sec:UT}

The short-range part of the NN force at N$^3$LO can be parametrized in terms of $15$ contact interactions $\sim \mathcal{O} (Q^4)$, which can be constructed out of the relative momenta of the nucleons \cite{Entem:2001cg,Epelbaum:2004fk}. By projecting these operators onto NN partial waves, one recognizes that $3$ out of $15$ operators are redundant in the sense that they cannot be determined from two-nucleon data. Indeed, the short-range interactions parametrizing the transitions $^1$S$_0 \, \to\, ^1$S$_0$, $^3$S$_1\,\to \, ^3$S$_1$ and $^3$S$_1\,\to \, ^3$D$_1$ are given by 
\begin{eqnarray}
\langle ^1{\rm S}_0, \, p' | \hat V_{\rm NN, \, short-range} | ^1{\rm S}_0, \, p \rangle  &=& \tilde
C_{^1{\rm S}_0} + C_{^1{\rm S}_0} (p^2 + p'^2) + D_{^1{\rm S}_0} p^2 p'^2 + D_{^1{\rm S}_0}^{\rm off} (p^2 -
                                                                           p'^2 )^2\,, \nonumber \\
 \langle ^3{\rm S}_1, \, p' | \hat V_{\rm NN, \, short-range} | ^3{\rm S}_1, \, p \rangle  &=& \tilde
C_{^3{\rm S}_1} + C_{^3{\rm S}_1} (p^2 + p'^2) + D_{^3{\rm S}_1} p^2 p'^2 + D_{^3{\rm S}_1}^{\rm off} (p^2 -
                                                                           p'^2 )^2\,, \nonumber \\ 
\langle ^3S_1 , \, p' | \hat V_{\rm NN, \, short-range} | ^3D_1 , \, p \rangle  &=&
C_{\epsilon_1} p^2 + D_{\epsilon_1} p^2 p'^2 + D_{\epsilon_1}^{\rm off} p^2(p'^2 - p^2)
\,, 
\end{eqnarray}
where $p$ and $p'$ denote the magnitudes of the initial and final momenta of the nucleons in the center-of-mass system, $\hat V$ signifies that this quantity is to be understood as an operator in momentum space (rather than the corresponding matrix elements denoted by $V$) and $\tilde
C_{a}$, $C_{a}$, $D_{a}$ and $D_{a}^{\rm off}$ refer to the corresponding LECs in the spectroscopic basis. The LECs  $D_{a}$ and $D_{a}^{\rm off}$ are given by linear combinations of the LECs 
$D_{1, \ldots , 15}$ accompanying the  order-$Q^4$ NN contact interactions, whose explicit form can be found in Ref.~\cite{Reinert:2017usi}. In the on-shell kinematics with $p' = p$, the contributions involving the LECs $D_a^{\rm off}$ vanish, signalling that these interactions are redundant. Indeed, they can be eliminated from the NN Hamiltonian by performing a unitary transformation (UT)
\beq
\hat U = \exp \Big(- \sum_i \beta_i \hat T_i\Big),
\eeq
where $\beta_i$ are real (dimensionful) rotation angles, while the anti-hermitian generators $\hat T_i$ are given by  \cite{Reinert:2017usi}
\beqa
   \langle \vec p_1^{\, \prime} \vec p_2^{\, \prime}|  \hat T_1 | \vec p_1 \vec p_2 \rangle &=& \vec{k} \cdot \vec{q} , \nn
     \langle \vec p_1^{\, \prime} \vec p_2^{\, \prime}|  \hat T_2 | \vec p_1 \vec p_2 \rangle &=& \vec{k} \cdot \vec{q} \; \vec{\sigma}_1 \cdot \vec{\sigma}_2 ,\nn
    \langle \vec p_1^{\, \prime} \vec p_2^{\, \prime}|  \hat T_3 | \vec p_1 \vec p_2 \rangle &=& \vec{\sigma}_1 \cdot \vec{k} \; \vec{\sigma}_2 \cdot \vec{q} + \vec{\sigma}_1 \cdot \vec{q} \; \vec{\sigma}_2 \cdot \vec{k}. 
\eeqa
Here, $\vec{q}, \vec{k}$ are given by the nucleon incoming (outgoing) momenta $\vec p_i$ ($\vec p_i^{\, \prime}$) via $\vec{q} = (\vec p_1^{\, \prime} - \vec p_2^{\, \prime}  - \vec p_1 + \vec p_2)/2$ and $\vec{k} = (\vec p_1^{\, \prime} - \vec p_2^{\, \prime}  + \vec p_1 - \vec p_2)/4$. According to the chiral power counting, the dominant effect of the UT $\hat U$ amounts to
\beq
\label{InducedTerms}
\delta \hat H = \hat U^\dagger H^{(0)} \hat U - \hat H^{(0)} \simeq \sum_i \big[ \big( \hat H_{\rm kin}^{(0)} + \hat V_{\rm short-range}^{(0)} +\hat H_{1\pi}^{(0)} \big), \; \beta_i \hat T_i \big]\,,
\eeq
where the superscript $(0)$ signifies that we only consider the leading-order (i.e., order-$Q^0$) contribution to the Hamiltonian.
It is straightforward to verify that two-nucleon operators stemming from the commutator with $\hat H_{\rm kin}^{(0)}$ have the same momentum dependence as the redundant short-range operators. Accordingly, it is possible to eliminate such operators from the NN potential by performing a UT with suitably chosen parameters $\beta_i$. Assuming that the LECs $D_i$ scale according to naive dimensional analysis as $|D_i | \sim F_\pi^{-2}\Lambda_b^{-4}$ \cite{Epelbaum:2014efa,Reinert:2017usi}, the unitary phases need to be chosen as $\beta_i = \mathcal{O} (m F_\pi^{-2} \Lambda_b^{-4})$. Removing the redundant off-shell structures by, e.g., setting $D_{^1{\rm S}_0}^{\rm off} = D_{^3{\rm S}_1}^{\rm off} = D_{\epsilon_1}^{\rm off} =0$ as done in Ref.~\cite{Reinert:2017usi} then implies that certain linear combinations of the subleading contact and $1\pi$-contact 3NFs, induced by $\hat V_{\rm short-range}^{(0)}$ and $\hat H_{1\pi}^{(0)}$ in Eq.~(\ref{InducedTerms}), are enhanced in the counting scheme with $m \sim \Lambda_b^2/M_\pi$ and should be included at N$^3$LO rather than N$^4$LO, see Ref.~\cite{Girlanda:2020pqn} for a discussion.\footnote{The considered unitary ambiguities reflect the well-known scheme dependence of 3NFs \cite{Polyzou:1990hks}, see also Ref.~\cite{Epelbaum:2025aan} for recent discussion.} 

In Ref.~\cite{Girlanda:2020pqn}, it was found that the most general form of the anti-hermitian short-range generators $\hat T_i$ involves two additional operators, which depend on the two-nucleon center-of-mass momentum  $\vec{P} = \vec{p}_1 + \vec{p}_2$,
\beqa
\langle \vec p_1^{\, \prime} \vec p_2^{\, \prime}|  \hat T_4 | \vec p_1 \vec p_2 \rangle &=& i \vec{P} \times \vec{k} \cdot (\vec{\sigma}_1 -\vec{\sigma}_2 ) , \nn
    \langle \vec p_1^{\, \prime} \vec p_2^{\, \prime}|  \hat T_5 | \vec p_1 \vec p_2 \rangle &=& \frac{1}{2} (\vec{\sigma}_1 \cdot \vec{P} \; \vec{\sigma}_2 \cdot \vec{q} - \vec{\sigma}_1 \cdot \vec{q} \; \vec{\sigma}_2 \cdot \vec{P}) . 
\eeqa
These two structures are shown in Ref.~\cite{Girlanda:2020pqn} to be related to the ``intrinsic'' non-minimal contributions to the boost operator and give rise to the additional order-$Q^4$ total-momentum-dependent NN contact interactions $\propto D_{16, 17}$. Such contact terms also vanish in the on-shell kinematics and can be eliminated from the Hamiltonian via suitably chosen UTs driven by the generators $\hat T_{4,5}$.  
Notice that the most general structure of the NN contact interactions with four momenta involves $\vec P$-dependent operators, whose coefficients $\propto m^{-4}$ and $m^{-2}$ are fixed in terms of the order-$Q^0$ (LO) and order-$Q^2$ (NLO) LECs $C_{S,T}$ and $C_{1, \ldots , 7}$, respectively,   as a consequence of the Poincar\'e symmetry, see Ref.~\cite{Filandri:2024ary} for explicit expressions. These relativistic corrections can be obtained by boosting the $\vec P$-independent contact interactions at LO and NLO using the minimal form of the boost operator. The appearance of $\vec P$-dependent NN contact interactions $\propto D_{16, 17}$, which survive in the static limit of $m \to \infty$, requires two-body contributions to the intrinsic boost operator of the order $\mathcal{O}(m^2)$ \cite{Girlanda:2020pqn}. In our opinion, this is incompatible with the assumed existence of a meaningful nonrelativistic limit for interacting nucleons, see Refs.~\cite{Forest:1995sg,Nasoni:2023adf} for a related discussion.
By requiring the absence of terms with positive powers of the nucleon mass in the two-body part of the boost operator,
$\vec P$-dependent NN contact interactions with coefficients unconstrained by the  Poincar\'e symmetry get suppressed by at least two inverse powers of the nucleon mass. Stated differently, we expect the corresponding LECs $D_{16, 17}$ to be of the order of $D_{16, 17} \sim F_\pi ^{-2} m ^{-2} \Lambda_b^{-2}$ rather than  $D_{16, 17} \sim F_\pi^{-2}\Lambda_b^{-4}$ as assumed in Ref.~\cite{Girlanda:2020pqn}. In the counting scheme for the nucleon mass  we employ, such terms start contributing at N$^5$LO (i.e., $\mathcal{O} (Q^6)$).  

The contributions to the $1\pi$-contact 3NFs induced by the UTs with the generators $\hat T_{1, \ldots , 5}$ can be obtained by calculating the commutator with $\hat H_{1\pi}^{(0)}$ in Eq.~(\ref{InducedTerms}), leading to\footnote{Our result does not fully agree with the corresponding expression in Ref.~\cite{Girlanda:2020pqn}.} 
\beqa
\label{InducedUT}
    \delta V &=& - \frac{g_A^2}{8 F_\pi^2}  \sum_{i \neq j \neq k} \frac{\vec{q}_k \cdot \vec{\sigma}_k \pmb{\tau}_i \cdot \pmb{\tau}_k }{q_k^2 + M_\pi^2} \Big\{  \beta_1 \vec{q}_j \cdot \vec{q}_k \; \vec{q}_k \cdot \vec{\sigma}_i  \nn
    &+& \beta_2 \big[ \vec{q}_j \cdot \vec{q}_k \; \vec{q}_k \cdot \vec{\sigma}_j + 2 i (\vec{k_i} - \vec{k}_j) \cdot \vec{q}_j \; \vec{q}_k \cdot \vec{\sigma}_i \times \vec{\sigma}_j \big] \nn
    &+& \beta_3 \big[ q_k^2 \vec{q}_j \cdot \vec{\sigma}_j +    \vec{q}_j \cdot \vec{q}_k \; \vec{q}_k \cdot \vec{\sigma}_j + 2 i \vec{q}_j \cdot \vec{\sigma}_j \; (\vec{k}_i - \vec{k}_j) \cdot \vec{q}_k \times \vec{\sigma}_i + 2 i (\vec{k}_i - \vec{k}_j)\cdot \vec{\sigma}_j \; \vec{q}_j \cdot \vec{q}_k \times \vec{\sigma}_i \big] \nn
    &+& 2 \beta_4 \big[ 2 \vec{k}_i \cdot \vec{q}_k \; \vec{k}_j \cdot \vec{\sigma}_i - 2 \vec{k}_j \cdot \vec{q}_k \; \vec{k}_i \cdot \vec{\sigma}_i - i \vec{q}_k \cdot \vec{\sigma}_i \vec{k}_j \cdot \vec{q}_k \times \vec{\sigma}_j \big] \nn
    &+& \beta_5 \big[  \vec{q}_j \cdot \vec{q}_k \; \vec{q}_k \cdot \vec{\sigma}_j - q_k^2 \vec{q}_j \cdot \vec{\sigma}_j - 2 i \vec{q}_j \cdot \vec{\sigma}_j \; (\vec{k}_i + \vec{k}_j) \cdot \vec{q}_k \times \vec{\sigma}_i + 2 i (\vec{k}_i + \vec{k}_j)\cdot \vec{\sigma}_j \; \vec{q}_j \cdot \vec{q}_k \times \vec{\sigma}_i \big]  \Big\} .
\eeqa
% In our opinion, Girlandas terms associated with $\gamma_4$ can not be correct. Moreover, for the terms associated with $\gamma_5$, we obtained for some contributions a different sign.
As explained above, switching off the off-shell contact interactions in the NN potential as done in Ref.~\cite{Reinert:2017usi} requires a promotion of the 3NFs, driven by the parameters $\beta_{1,2,3}$, from N$^4$LO to N$^3$LO. By matching the corresponding antisymmetrized expressions in Eqs.~(\ref{InducedUT}) and (\ref{3NFDlike}), we read out the induced contributions to the LECs $F_i$: 
\beqa
\label{InducedLECsUT}
F_{1}^{\rm UT} &=&  F_{5}^{\rm UT} = F_{6}^{\rm UT} = F_{7}^{\rm UT} = F_{8}^{\rm UT} = F_{9}^{\rm UT} = F_{10}^{\rm UT} = F_{11}^{\rm UT} = 0, \nn
F_{2}^{\rm UT} &=&  F_{4}^{\rm UT}  = 4 F_{12}^{\rm UT}  = 4 F_{15}^{\rm UT}  = - 2 F_{16}^{\rm UT}  =  \beta_3,\nn
F_{3}^{\rm UT} &=& - 2 (\beta_2 + \beta_3), \nn
F_{13}^{\rm UT} &=& \frac{1}{4} (\beta_1 + \beta_2 +\beta_3), \nn
F_{14}^{\rm UT} &=& \frac{1}{4} (\beta_1 + \beta_2 ).
\eeqa
As already mentioned above, these  3NFs formally appear already at N$^3$LO given 
that $| \beta_{1,2,3}| \sim m F_\pi^{-2} \Lambda_b^{-4}$.
Notice that the UTs $\propto \beta_{1,2,3}$ also induce a structure corresponding to the operator $\fett O_{17}$, which leads to (formally enhanced) shifts of the N$^2$LO LECs $D$ and $E$ but has no practical implications. Finally, we emphasize that the 3NFs in Eq.~(\ref{InducedUT}) induced by $\beta_{4,5}$ can, as one may expect, not be absorbed into shifts of the LECs $F_i$. As explained  above, we expect $ \beta_{4,5}$ to scale according to $| \beta_{4,5}| \sim m^{-1} F_\pi ^{-2} \Lambda_b^{-2}$, so that the corresponding induced 3NF appear at N$^5$LO.

\section{Resonance saturation}
\label{sec:Delta}

Up to now, our discussion was based on the $\Delta$-less formulation of chiral perturbation theory, i.e.~we have only taken into account nucleons and pions as active degrees of freedom. On the other hand, in the $\Delta$-full formulation based on the small scale expansion scheme with $\Delta \equiv m_\Delta - m \sim M_\pi = \mathcal{O} (\epsilon )$ \cite{Hemmert:1997ye}, the first tree-level correction to the $1\pi$-contact 3NF topology appears already at N$^3$LO from diagrams involving an intermediate $\Delta$-excitation. This suggests that certain linear combinations of the LECs $F_i$ in the  $\Delta$-less framework are enhanced due to the small excitation energy of the $\Delta$-isobar. Assuming that the values of the LECs $F_i$ are saturated by the $\Delta$-isobar, the dominant contributions to the considered \textit{D}-like N$^4$LO 3NF can be approximated in terms of a fewer number of LECs, which parametrize the most general form of the NN$\to$N$\Delta$ interaction with two derivatives. Notice that the LECs $D$ and $E$ of the N$^2$LO 3NF are not saturated by the $\Delta$ since the NN$\to$N$\Delta$ vertex without derivatives vanishes for antisymmetrized NN states \cite{Epelbaum:2005pn,Epelbaum:2007sq}. 

To derive the constraints on the $F_i$'s from resonance saturation we follow the approach of sec.~\ref{sec:Method1} and seek for the most general parametrization of the operators
$\Tilde{\fett O}{}_i^\Delta \sim \mathcal{O} (Q^3)$,
which involve two spin and two isospin $1/2 \to 3/2$ transition matrices
$\vec {\mathcal{S}}$ and $\fett {\mathcal{T}}$. These $2 \times 4$ matrices  fulfill the relationships $\mathcal{S}^i (\mathcal{S}^\dagger)^j = (2 \delta^{ij} - i \epsilon^{ijk} \sigma^k)/3$ and $\mathcal{T}^a (\mathcal{T}^\dagger)^b = (2 \delta^{ab} - i \epsilon^{ijk} \tau^k)/3$ \cite{Ericson:1988gk}. 
The operators $\Tilde{\fett O}{}_i^\Delta$ we consider incorporate both the NN$\to$N$\Delta$ and $\Delta \to$N$\pi$ vertex structures, so that the corresponding 3NF potentials are given by the expression in Eq.~(\ref{PotFromO}) with $\Tilde{\fett O}_i$ replaced by $\Tilde{\fett O}{}_i^\Delta$.
Notice that in the small-scale expansion, these operators scale as $\Tilde{\fett O}{}_i^\Delta \sim \mathcal{O} (\epsilon^2)$.
We found $5$ possible structures for the operators $\Tilde{\fett O}{}_i^\Delta$:
\beqa
\Tilde{\fett O}{}_1^{\Delta} &=& \big[i \; \Vec{q_3} \cdot \Vec{\mathcal{S}_2} \Vec{\mathcal{S}_2^\dagger} \cdot ((\Vec{q}-\Vec{q_3}/2 ) \times (\Vec{k}-\Vec{q_3}/4) )  \; \boldsymbol{\mathcal{T}}_2 \boldsymbol{\mathcal{T}}_2^\dagger \cdot \boldsymbol{\tau}_1 \; +\;  1 \leftrightarrow 2\big] \; - \; \mbox{h.c.}, \nn 
\Tilde{\fett O}{}_2^\Delta &=&  \big[\Vec{q_3} \cdot \Vec{\mathcal{S}_2} \Vec{\mathcal{S}_2^\dagger} \cdot \Vec{\sigma_1} (\Vec{q}-\Vec{q_3}/2 )^2 \; \boldsymbol{\mathcal{T}}_2 \boldsymbol{\mathcal{T}}_2^\dagger \cdot \boldsymbol{\tau}_1 \; + \; 1 \leftrightarrow 2\big] \; - \; \mbox{h.c.}, \nn 
\Tilde{\fett O}{}_3^\Delta &=&  \big[ \Vec{q_3} \cdot \Vec{\mathcal{S}_2} \Vec{\mathcal{S}_2^\dagger} \cdot (\Vec{q}-\Vec{q_3}/2 ) \Vec{\sigma_1} \cdot(\Vec{q}-\Vec{q_3}/2) \; \boldsymbol{\mathcal{T}}_2 \boldsymbol{\mathcal{T}}_2^\dagger \cdot \boldsymbol{\tau}_1 \; +\;  1 \leftrightarrow 2\big] \; -\;  \mbox{h.c.}, \nn
\Tilde{\fett O}{}_4^\Delta  &=&  \big[ \Vec{q_3} \cdot \Vec{\mathcal{S}_2} \Vec{\mathcal{S}_2^\dagger} \cdot (\Vec{k}-\Vec{q_3}/4 ) \Vec{\sigma_1} \cdot(\Vec{k}-\Vec{q_3}/4) \; \boldsymbol{\mathcal{T}}_2 \boldsymbol{\mathcal{T}}_2^\dagger \cdot \boldsymbol{\tau}_1 \; +\;  1 \leftrightarrow 2\big] \; -\;  \mbox{h.c.}, \nn  
\Tilde{\fett O}{}_5^\Delta &=&  \big[\Vec{q_3} \cdot \Vec{\mathcal{S}_2} \Vec{\mathcal{S}_2^\dagger} \cdot \Vec{\sigma_1} (\Vec{k}-\Vec{q_3}/4 )^2 \; \boldsymbol{\mathcal{T}}_2 \boldsymbol{\mathcal{T}}_2^\dagger \cdot \boldsymbol{\tau}_1 \; +\;  1 \leftrightarrow 2\big]\;  -\;  \mbox{h.c.}.
\eeqa
After antisymmetrization with respect to the initial nucleons $1$ and $2$, one observes that  the operator $\Tilde{\fett O}{}_5^\Delta $ is redundant since ${\fett O}{}_5^\Delta = -(1/4)  {\fett O}{}_2^\Delta$. The leading contributions of the delta resonance to the $1\pi$-contact 3NF thus originate from the operators  ${\fett O}{}_{1, \ldots , 4}^\Delta$. By parametrizing the most general operator structure via $\sum_{i=1}^4 \alpha_i  {\fett O}{}_i^\Delta $ and matching the resulting 3NF fo the expression in Eq.~(\ref{3NFDlike}), we read out the delta contributions to the LECs $F_i$ as follows:
\beqa
\label{LECsDelta}
F_1^\Delta &=& -4  F_2^\Delta  = 4 F_4^\Delta = \frac{8 \alpha_1}{9 \Delta},
\nn F_3^\Delta &=&  F_{11}^\Delta = 0,
\nn  F_5^\Delta &=& - \frac{4 \alpha_2 + 2 \alpha_3}{9 \Delta}, 
\nn F_6^\Delta &=& \frac{2 \alpha_3 + \alpha_4}{9 \Delta}, 
\nn F_7^\Delta &=& \frac{8 \alpha_2 - \alpha_4}{18 \Delta}, 
\nn F_8^\Delta &=& \frac{4 \alpha_2}{9 \Delta}, 
\nn F_9^\Delta &=& \frac{8 \alpha_3 + \alpha_4}{18 \Delta}, 
\nn F_{10}^\Delta &=& 4  F_{12}^\Delta  = \frac{4 \alpha_3 - \alpha_4}{9 \Delta}, 
\nn F_{13}^\Delta &=& \frac{2 \alpha_1 - 8 \alpha_2 - 4 \alpha_3 + \alpha_4}{18 \Delta},
\nn F_{14}^\Delta &=& \frac{-2 \alpha_1 - 16 \alpha_2 - 8 \alpha_3 + \alpha_4}{36 \Delta}, 
\nn F_{15}^\Delta &=& \frac{-2 \alpha_1  - 4 \alpha_3 + \alpha_4}{18 \Delta}, 
\nn F_{16}^\Delta &=& \frac{2 \alpha_1  - 8 \alpha_3 - \alpha_4}{36 \Delta},
\eeqa
where the constants $\alpha_i$ are expected to be of the order of $|\alpha_i | \sim 1/ (F_\pi^2 \Lambda_b^2)$.   We have verified  that the contributions to the LECs $F_i$, parametrized by $\alpha_{1, \ldots , 4}$ and $\beta_{1, 2, 3}$, are independent of each other. This can be expected given that the UT-induced contributions to the 3NF considered in sec.~\ref{sec:UT} originate from reducible-like diagrams, while those driven by intermediate $\Delta$-excitations correspond to three-nucleon-irreducible diagrams.

%*********************************************************%
%-----------------------Summary---------------------------%
%*********************************************************%
\section{Summary}
\label{sec:summary}

In this paper we have analyzed the subleading corrections to the one-pion-exchange-contact 3NF. Our main results can be summarized as follows.
\begin{itemize}
\item
  We have shown that the most general expression for the $1\pi$-contact 3NF at order $Q^5$ (i.e., at N$^4$LO) can be parametrized in terms of $16$ unknown LECs $F_{1, \ldots , 16}$.
 % \footnote{This result is, strictly speaking, valid under the restriction that two-body contributions to the boost operator remain finite in the $m \to \infty$ limit.}
The isospin-$3/2$ part of this 3NF is found to depend only on $6$ linear combinations of $F_i$'s. On the other hand, all $16$ LECs contribute to the isospin-$1/2$ channel and thus can, at least in principle, be determined from nucleon-deuteron scattering data.  
\item
  We have discussed the relationship between the off-shell part of the short-range NN potential and the $1\pi$-contact 3NF. The chiral EFT NN potentials of Refs.~\cite{Reinert:2017usi,Reinert:2020mcu} make a specific choice for the off-shell behavior by switching off the corresponding N$^3$LO contact interactions using $D_{^1{\rm S}_0}^{\rm off} = D_{^3{\rm S}_1}^{\rm off} = D_{\epsilon_1}^{\rm off}   =0$. Fixing these off-shell ambiguities has no impact on the calculated NN observables, but requires certain linear combinations of the subleading contact and $1\pi$-contact 3NFs to be promoted from N$^4$LO to N$^3$LO (for the employed counting of the nucleon mass). The resulting N$^3$LO contributions to the $1\pi$-contact 3NF can be expressed in terms of $3$ parameters $\beta_{1,2,3}$ as given in Eq.~(\ref{InducedLECsUT}).
\item
Assuming the validity of the resonance saturation hypothesis, one may expect the dominant contributions to the $1\pi$-contact 3NF to be driven by the intermediate $\Delta$(1232) excitation mechanism. Using the $\Delta$-full formulation of chiral EFT within the small-scale expansion scheme, we worked out the leading contributions of the delta-isobar to the considered type of 3NF, which can be reproduced in the $\Delta$-less framework by setting $F_i = F_i^\Delta$ as given in Eq.~(\ref{LECsDelta}). This allows one to approximate the dominant effects from the subleading $1\pi$-contact 3NF in terms of just $4$ unknown coefficients $\alpha_{1, \ldots , 4}$, which parametrize the short-range order-$\mathcal{O} (Q^2)$ NN$\, \to \,$N$\Delta$ interactions.
\end{itemize}
Our work constitutes an important step towards the ongoing quest to establish high-precision 3NFs in chiral EFT \cite{Endo:2024cbz}, which is expected to require pushing the chiral expansion to N$^4$LO. Our results show that the short-range structure of the 3NF at this expansion order is rather rich and by far not exhausted by the purely contact interactions parametrized in terms of the LECs $E_{1, \ldots , 13}$ \cite{Girlanda:2011fh}, which were shown to allow for a significantly improved description of three-nucleon scattering data \cite{Girlanda:2018xrw,Epelbaum:2019zqc,Witala:2022rzl,Girlanda:2023znc}. It remains to be seen whether the contributions of the subleading $1\pi$-contact 3NFs to nucleon-deuteron scattering observables can be disentangled from those of the $E_i$-terms, which is a prerequisite for their reliable determination from three-nucleon data. Alternative strategies for fixing the LECs $F_i$ could rely on using experimental data in $A > 3$ systems and/or employing controlled approximations like, e.g., the  resonance saturation hypothesis, in order to reduce the number of adjustable parameters. 
% Acknowledgement
\begin{acknowledgments}
  We are grateful to Ashot Gasparyan and Jambul Gegelia for useful discussions and to
  Arseniy Filin for his help with implementing the calculations in {\it Mathematica}.  
This work has been supported in part by the European Research Council (ERC)
under the European Union’s Horizon 2020 research and innovation
programme (grant agreement No.~885150), by the MKW NRW under the
funding code NW21-024-A, by JST ERATO (Grant No. JPMJER2304) and by JSPS
KAKENHI (Grant No. JP20H05636) and by BMBF through the ErUM-Data project DEMOS.
\end{acknowledgments}

		%******************************************************************************************
		\appendix
\renewcommand{\theequation}{\thesection.\arabic{equation}}
                
		\section{Nonrelativistic reduction of the $\pi$NN Lagrangian}\label{Appendix_Principle Value Integrals}
\label{App}
                
In the static limit of $m \to \infty$, the expressions of the various operators in the covariant $\pi$NN Lagrangian at order $Q^3$ can be written in terms of the nonrelativistic operators defined in Table~\ref{Table3}  as follows: 
\beqa
\label{NRreduction}
O_1^{\RN{3}, \rm \, static} &=&-O_{14}^{\RN{2}}+O_{19}^{\RN{2}}-O_{22}^{\RN{2}}-O_{23}^{\RN{2}}+O_{26}^{\RN{2}}-O_{29}^{\RN{2}}-O_{30}^{\RN{2}}-\frac{O_{38}^{\RN{2}}}{2}+\frac{O_{40}^{\RN{2}}}{2}-\frac
{O_{42}^{\RN{2}}}{2}+\frac{O_{44}^{\RN{2}}}{2}+O_{62}^{\RN{2}},\nn
O_2^{\RN{3}, \rm \, static} &=&
O^{\RN{2}}_{14}-\frac{O^{\RN{2}}_{38}}{2}-\frac{O^{\RN{2}}_{40}}{2}-\frac{O^{\RN{2}}_{42}}{2}-\frac{O^{\RN{2}}_{44}}{2}-O^{\RN{2}}_{62}, \nn
O_3^{\RN{3}, \rm \, static}&=&
O ^{\RN{2}}_{14}-\frac{O ^{\RN{2}}_{38}}{2}-\frac{O ^{\RN{2}}_{40}}{2}-\frac{O ^{\RN{2}}_{42}}{2}-\frac{O ^{\RN{2}}_{44}}{2}+O ^{\RN{2}}_{62}-O ^{\RN{2}}_{65},\nn
O_4^{\RN{3}, \rm \, static}&=&-O ^{\RN{2}}_{14}-O ^{\RN{2}}_{19}+O ^{\RN{2}}_{22}+O ^{\RN{2}}_{23}-O ^{\RN{2}}_{26}+O ^{\RN{2}}_{29}+O ^{\RN{2}}_{30}+\frac{O ^{\RN{2}}_{38}}{2}-\frac{O ^{\RN{2}}_{40}}{2}+\frac
{O ^{\RN{2}}_{42}}{2}-\frac{O ^{\RN{2}}_{44}}{2}-O ^{\RN{2}}_{62}+O ^{\RN{2}}_{65}, \nn
O_5^{\RN{3}, \rm \, static}&=&
O ^{\RN{2}}_4+O ^{\RN{2}}_{38}+\frac{O ^{\RN{2}}_{40}}{2}+\frac{O ^{\RN{2}}_{42}}{2}+O ^{\RN{2}}_{44}+O ^{\RN{2}}_{62}-\frac{O ^{\RN{2}}_{65}}{2}, \nn
O^{\RN{3}, \rm \, static}_6&=&
-O^{\RN{2}}_4-O^{\RN{2}}_{19}+O^{\RN{2}}_{22}+O^{\RN{2}}_{23}-O^{\RN{2}}_{26}+O^{\RN{2}}_{29}+O^{\RN{2}}_{30}+\frac{O^{\RN{2}}_{38}}{2}+O^{\RN{2}}_{42}-\frac{O^{\RN{2}}_{44}}{2}-O
   _{62}+\frac{O^{\RN{2}}_{65}}{2}, \nn
O^{\RN{3}, \rm \, static}_7&=&
-O^{\RN{2}}_5-O^{\RN{2}}_6-2 O^{\RN{2}}_{39}-O^{\RN{2}}_{41}-O^{\RN{2}}_{43}-2 O^{\RN{2}}_{45}, \nn
O^{\RN{3}, \rm \, static}_8&=&
O^{\RN{2}}_5+O^{\RN{2}}_6-O^{\RN{2}}_{41}-O^{\RN{2}}_{43}, \nn
O^{\RN{3}, \rm \, static}_{11}&=&
-O^{\RN{2}}_{19}+O^{\RN{2}}_{22}+O^{\RN{2}}_{23}-O^{\RN{2}}_{26}+O^{\RN{2}}_{29}+O^{\RN{2}}_{30}+\frac{O^{\RN{2}}_{38}}{2}-\frac{O^{\RN{2}}_{40}}{2}+\frac{O^{\RN{2}}_{42}
   }{2}-\frac{O^{\RN{2}}_{44}}{2}-O^{\RN{2}}_{62}, \nn
O^{\RN{3}, \rm \, static}_{12}&=&
O^{\RN{2}}_{62}, \nn
O^{\RN{3}, \rm \, static}_{13}&=&
O^{\RN{2}}_{14}-\frac{O^{\RN{2}}_{19}}{2}-O^{\RN{2}}_{20}-\frac{3 O^{\RN{2}}_{22}}{2}-\frac{3
   O^{\RN{2}}_{23}}{2}-\frac{O^{\RN{2}}_{26}}{2}-O^{\RN{2}}_{27}+O^{\RN{2}}_{28}-\frac{3 O^{\RN{2}}_{29}}{2}-\frac{3
   O^{\RN{2}}_{30}}{2}-O^{\RN{2}}_{31} -\frac{3
   O^{\RN{2}}_{38}}{4}-\frac{O^{\RN{2}}_{39}}{2}-\frac{O^{\RN{2}}_{40}}{4}, \nn
 &+&\frac{O^{\RN{2}}_{41}}{2}-\frac{3
   O^{\RN{2}}_{42}}{4}-\frac{O^{\RN{2}}_{43}}{2}-\frac{O^{\RN{2}}_{44}}{4}+\frac{O^{\RN{2}}_{45}}{2}+\frac{3
   O^{\RN{2}}_{51}}{4}-O^{\RN{2}}_{52}+\frac{3 O^{\RN{2}}_{54}}{4}, \nn
O^{\RN{3}, \rm \, static}_{14}&=&
-O^{\RN{2}}_{14}-\frac{O^{\RN{2}}_{19}}{2}-O^{\RN{2}}_{20}-\frac{3
   O^{\RN{2}}_{22}}{2}+\frac{O^{\RN{2}}_{23}}{2}-\frac{O^{\RN{2}}_{26}}{2}-O^{\RN{2}}_{27}-O^{\RN{2}}_{28}-\frac{3
   O^{\RN{2}}_{29}}{2}+\frac{O^{\RN{2}}_{30}}{2}+O^{\RN{2}}_{31} -\frac{O^{\RN{2}}_{38}}{4}+\frac{O^{\RN{2}}_{39}}{2}+\frac{O^{\RN{2}}_{40}}{4}, \nn
 &-&
   \frac{O^{\RN{2}}_{41}}{2}-\frac{O^{\RN{2}}_{42}}{4}+\frac{O^{\RN{2}}_{43}}{2}+\frac{O^{\RN{2}}_{44}}{4}-\frac{O^{\RN{2}}_{45}}{2}-\frac{O^{\RN{2}}_{51}}{4}+O^{\RN{2}}_{52}-\frac{O^{\RN{2}}_{54}}{4}, \nn
O^{\RN{3}, \rm \, static}_{15}&=&
-2 O^{\RN{2}}_{19}+2 O^{\RN{2}}_{22}+3 O^{\RN{2}}_{23}-2 O^{\RN{2}}_{26}+2
   O^{\RN{2}}_{29}+O^{\RN{2}}_{30}+O^{\RN{2}}_{38}-O^{\RN{2}}_{40}+O^{\RN{2}}_{42}-O^{\RN{2}}_{44}+O^{\RN{2}}_{52}-O^{\RN{2}}_{53}-O^{\RN{2}}_{54}, \nn
O^{\RN{3}, \rm \, static}_{16}&=&
-O^{\RN{2}}_{23}+O^{\RN{2}}_{30}-O^{\RN{2}}_{52}+O^{\RN{2}}_{53}+O^{\RN{2}}_{54}, \nn
O^{\RN{3}, \rm \, static}_{19}&=&
O^{\RN{2}}_{15}-\frac{3 O^{\RN{2}}_{19}}{2}+\frac{3 O^{\RN{2}}_{22}}{2}+\frac{O^{\RN{2}}_{23}}{2}-\frac{3 O^{\RN{2}}_{26}}{2}+2
   O^{\RN{2}}_{28}+\frac{3 O^{\RN{2}}_{29}}{2}-\frac{3 O^{\RN{2}}_{30}}{2}-2 O^{\RN{2}}_{31}+\frac{3 O^{\RN{2}}_{38}}{4}-\frac{3
     O^{\RN{2}}_{39}}{2} -\frac{3 O^{\RN{2}}_{40}}{4}+\frac{O^{\RN{2}}_{41}}{2}, \nn
   &+&\frac{3 O^{\RN{2}}_{42}}{4}-\frac{3
   O^{\RN{2}}_{43}}{2}-\frac{3 O^{\RN{2}}_{44}}{4}+\frac{O^{\RN{2}}_{45}}{2}+\frac{5
   O^{\RN{2}}_{51}}{4}+\frac{O^{\RN{2}}_{54}}{4}+O^{\RN{2}}_{61}-O^{\RN{2}}_{62}, \nn
O^{\RN{3}, \rm \, static}_{20}&=&
-O^{\RN{2}}_{15}+2 O^{\RN{2}}_{19}-2 O^{\RN{2}}_{22}-O^{\RN{2}}_{23}+2 O^{\RN{2}}_{26}-2 O^{\RN{2}}_{28}-2 O^{\RN{2}}_{29}+O^{\RN{2}}_{30}+2
O^{\RN{2}}_{31}-O^{\RN{2}}_{38}+O^{\RN{2}}_{39}+O^{\RN{2}}_{40}-O^{\RN{2}}_{41}-O^{\RN{2}}_{42}, \nn
&+&O^{\RN{2}}_{43}+O^{\RN{2}}_{44}-O^{\RN{2}}_{45}+O^{\RN{2}}_{54}-O^{\RN{2}}_{61}+O^{\RN{2}}_{62}, \nn
O^{\RN{3}, \rm \, static}_{21}&=&
O^{\RN{2}}_{19}+O^{\RN{2}}_{29}, \nn
O^{\RN{3}, \rm \, static}_{22}&=&
O^{\RN{2}}_{22}+O^{\RN{2}}_{26}, \nn
O^{\RN{3}, \rm \, static}_{23}&=&
\frac{O^{\RN{2}}_{51}}{2}+O^{\RN{2}}_{52}+O^{\RN{2}}_{53}+\frac{O^{\RN{2}}_{54}}{2}-O^{\RN{2}}_{62}, \nn
O^{\RN{3}, \rm \, static}_{24}&=&
O^{\RN{2}}_{19}-O^{\RN{2}}_{22}-O^{\RN{2}}_{23}+O^{\RN{2}}_{26}-O^{\RN{2}}_{29}-O^{\RN{2}}_{30}-\frac{O^{\RN{2}}_{38}}{2}+\frac{O^{\RN{2}}_{40}}{2}-\frac{O^{\RN{2}}_{42}}
   {2}+\frac{O^{\RN{2}}_{44}}{2}+\frac{O^{\RN{2}}_{51}}{2}-O^{\RN{2}}_{52}-O^{\RN{2}}_{53}+\frac{O^{\RN{2}}_{54}}{2}+O^{\RN{2}}_{62}, \nn
O^{\RN{3}, \rm \, static}_{25}&=&
-2 O^{\RN{2}}_{19}+2 O^{\RN{2}}_{22}+2 O^{\RN{2}}_{23}-2 O^{\RN{2}}_{26}+2 O^{\RN{2}}_{29}+2 O^{\RN{2}}_{30}+O^{\RN{2}}_{38}-O^{\RN{2}}_{40}+O^{\RN{2}}_{42}-O^{\RN{2}}_{44}, \nn
O^{\RN{3}, \rm \, static}_{26}&=&
-2 O^{\RN{2}}_{19}+2 O^{\RN{2}}_{22}+2 O^{\RN{2}}_{23}-2 O^{\RN{2}}_{26}+2 O^{\RN{2}}_{29}+2 O^{\RN{2}}_{30}+O^{\RN{2}}_{38}-O^{\RN{2}}_{40}+O^{\RN{2}}_{42}-O^{\RN{2}}_{44}, \nn
O^{\RN{3}, \rm \, static}_{27}&=&
O^{\RN{2}}_{15}-O^{\RN{2}}_{19}+O^{\RN{2}}_{22}+O^{\RN{2}}_{23}-O^{\RN{2}}_{26}+O^{\RN{2}}_{29}+O^{\RN{2}}_{30}+\frac{O^{\RN{2}}_{38}}{2}-\frac{O^{\RN{2}}_{40}}{2}+\frac{
   O^{\RN{2}}_{42}}{2}-\frac{O^{\RN{2}}_{44}}{2}-O^{\RN{2}}_{62}, \nn
O^{\RN{3}, \rm \, static}_{28}&=&
-O^{\RN{2}}_{15}+\frac{O^{\RN{2}}_{19}}{2}-\frac{O^{\RN{2}}_{22}}{2}-\frac{O^{\RN{2}}_{23}}{2}+\frac{O^{\RN{2}}_{26}}{2}-\frac{O^{\RN{2}}_{29}}
   {2}-\frac{O^{\RN{2}}_{30}}{2}-\frac{O^{\RN{2}}_{38}}{4}+\frac{O^{\RN{2}}_{39}}{2}+\frac{O^{\RN{2}}_{40}}{4}+\frac{O^{\RN{2}}_{41}}{
     2}-\frac{O^{\RN{2}}_{42}}{4}+\frac{O^{\RN{2}}_{43}}{2}+\frac{O^{\RN{2}}_{44}}{4}, \nn
   &+&\frac{O^{\RN{2}}_{45}}{2}+\frac{O^{\RN{2}}_{51}}{4
   }+\frac{O^{\RN{2}}_{54}}{4}+O^{\RN{2}}_{62}, \nn
O^{\RN{3}, \rm \, static}_{29}&=&
O^{\RN{2}}_{15}-\frac{O^{\RN{2}}_{19}}{2}+\frac{O^{\RN{2}}_{22}}{2}+\frac{O^{\RN{2}}_{23}}{2}-\frac{O^{\RN{2}}_{26}}{2}+\frac{O^{\RN{2}}_{29}}{
   2}+\frac{O^{\RN{2}}_{30}}{2}+\frac{O^{\RN{2}}_{38}}{4}-\frac{O^{\RN{2}}_{39}}{2}-\frac{O^{\RN{2}}_{40}}{4}-\frac{O^{\RN{2}}_{41}}{2
 }+\frac{O^{\RN{2}}_{42}}{4}-\frac{O^{\RN{2}}_{43}}{2}-\frac{O^{\RN{2}}_{44}}{4}, \nn
 &-&\frac{O^{\RN{2}}_{45}}{2}-\frac{O^{\RN{2}}_{51}}{4}
   -\frac{O^{\RN{2}}_{54}}{4}+O^{\RN{2}}_{61}-O^{\RN{2}}_{65}, \nn
O^{\RN{3}, \rm \, static}_{30}&=&
-O^{\RN{2}}_{15}+\frac{O^{\RN{2}}_{51}}{2}+\frac{O^{\RN{2}}_{54}}{2}-O^{\RN{2}}_{61}+O^{\RN{2}}_{65}. 
   \eeqa

\bibliography{references}
\end{document}